\documentclass[review]{elsarticle}

\usepackage{lineno,hyperref}
\modulolinenumbers[5]

\journal{Journal of Computational Physics}

\usepackage{graphicx}
\usepackage{amsmath,amssymb}
\usepackage{xcolor}
\usepackage{mathrsfs}
\usepackage{paralist}
\usepackage{multirow}
\usepackage[T1]{fontenc}
\usepackage[utf8]{inputenc}
\usepackage[english]{babel}
\usepackage{subfigure}
\DeclareMathAlphabet\mathbfcal{OMS}{cmsy}{b}{n}
\usepackage{caption}
\usepackage{lineno}
\usepackage{bbm}
\usepackage[normalem]{ulem}
\usepackage{float}

\newcommand{\ud}{\mathrm{d}}

\newcommand{\ie}{\emph{i.e.}} 
\newcommand{\Ma}{\mathrm{Ma}} 

\newcommand{\vect}[1]{\boldsymbol{\mathrm{#1}}}

\newcommand{\pd}[2]{\frac{\partial{#1}}{\partial{#2}}} 
\newcommand{\der}[2]{\frac{\ud{#1}}{\ud{#2}}} 

\newcommand{\del}[1]{}

\usepackage[a4paper,bindingoffset=0.0in,%
            left=0.8in,right=0.8in,top=1.in,bottom=1.in,%
            footskip=.4in]{geometry}

\begin{document}

\begin{frontmatter}

\title{Fully-discrete spatial eigenanalysis of discontinuous spectral element methods: insights into well-resolved and under-resolved vortical flows}

\author[INSA]{Niccol{\`o} Tonicello\corref{mycorrespondingauthor}}
\cortext[mycorrespondingauthor]{Corresponding author}
\ead{niccolo.tonicello@gmail.com}
\author[ITA]{Rodrigo C. Moura}
\author[INSA]{Guido Lodato}
\author[NUS]{Gianmarco Mengaldo}

\address[INSA]{Normandie Universit{\'e}, INSA et Universit{\'e} de Rouen, St.~Etienne du Rouvray (Rouen), France} 
\address[ITA]{Instituto Tecnológico de Aeronáutica, São José dos Campos, Brazil}
\address[NUS]{National University of Singapore, Singapore} 

\begin{abstract}
This study presents a comprehensive spatial eigenanalysis of fully-discrete discontinuous spectral element methods, now generalizing previous spatial eigenanalysis that did not include time integration errors.
The influence of discrete time integration is discussed in detail for different explicit Runge-Kutta (1st to 4th order accurate) schemes combined with either Discontinuous Galerkin (DG) or Spectral Difference (SD) methods, both here recovered from the Flux Reconstruction (FR) scheme.
Selected numerical experiments using the improved SD method by Liang and Jameson~\cite{liang2009high,liang2009spectral,jameson:10} are performed to quantify the influence of time integration errors on actual simulations. 
These involve test cases of varied complexity, from one-dimensional linear advection equation studies to well-resolved and under-resolved inviscid vortical flows. 
The effect of mesh regularity is also considered, where time integration errors are found to be, in the case of irregular grids, less pronounced than those of the spatial discretisation. Still, when simulations are well-resolved, the overall order of accuracy of the (fully-discrete) method of choice is limited to that of the time integration scheme. Moreover, it is shown that, while both well-resolved and under-resolved simulations of linear problems correlate well with the eigenanalysis prediction of time integration errors, the correlation can be much worse for under-resolved nonlinear problems.
In fact, for the under-resolved vortical flows considered, the predominance of spatial errors made it practically impossible for time integration errors to be distinctly identified. Nevertheless, for well-resolved nonlinear simulations, the effect of time integration errors could still be recognized. As a result, the eigenanalysis predictions are expected to hold (even if partially) in direct numerical simulations of turbulence. This highlights that the interaction between space and time discretisation errors is more complex than otherwise anticipated, contributing to the current understanding about when eigenanalysis can effectively predict the behavior of numerical errors in practical under-resolved nonlinear problems, including under-resolved turbulence computations.
\end{abstract}


\end{frontmatter}

\section{Introduction}
Along with the increasing computational power experienced in the last few decades, the use of Computational Fluid Dynamics (CFD) is becoming a more and more common tool in industrial aerodynamic analysis and design~\cite{slotnick2014cfd}.
The expanding capabilities of modern computers have fuelled the development of a large number of innovative numerical high-order schemes for fluid dynamics, whose accuracy clearly surpasses that of more popular low order methods.
In particular, much interest has been focused on spectral element methods due to their geometric flexibility and promising computational efficiency~\cite{vincent2011facilitating}. These include the Discontinuous Galerkin (DG)~\cite{hesthaven2007nodal,cockburn:98,cockburn:98b}, Spectral Difference (SD)~\cite{kopriva1996conservative,liu:06a,wang:07} and Flux Reconstruction (FR)~\cite{huynh2007flux,vincent2011new} schemes. The use of high-order numerical schemes has shown promising results in the simulation of turbulent flows, in both Direct Numerical Simulation (DNS)~\cite{chapelier2014evaluation,de2018use,bassi2016development,krank2018direct} and Large-Eddy Simulations (LES)~\cite{lodato2021large,tonicello2021analysis,fernandez2018ability,moura2017eddy,de2019performance,mengaldo2021industry}. Due to the rapidly developing field of high-order methods, especially in fluid dynamics, the research community is constantly looking for more efficient and reliable tools to quantify the peculiar numerical properties of such schemes. Indeed, for under-resolved simulations (such as LES), the role played by the numerical scheme can be of fundamental importance. For instance, the same computation, performed with different methods, can lead to dramatically different outcomes in realistic flow configurations~\cite{cox2021accuracy,moura2017eddy,fernandez2018ability}.  

Spectral analyses based on the numerical discretisation of the linear advection equation are a common tool in the evaluation of the intrinsic numerical properties of high-order methods for advection dominated flows.
Such techniques are particularly simple to implement and computationally inexpensive.
However simplistic, they are capable of providing useful information regarding the scheme, such as numerical dispersion and diffusion.
The most widely used spectral analysis technique is commonly referred to as \emph{temporal eigenanalysis}~\cite{lele1992compact,bogey2004family,van2008stability,vincent2011insights,moura2015linear,vanharen2017revisiting}, which quantifies how specific spatial oscillations will evolve over time under periodic conditions in space.
The complementary analysis technique, called \emph{spatial eigenanalysis}~\cite{hu2002eigensolution,mengaldo2018spatial,mengaldo2018spatial2,moura2020spatial,tonicello2021comparative}, considers instead how wave-like solutions of given frequency behave while propagating in space under inflow-outflow conditions.
Even if similar and intimately connected from a theoretical perspective, these two approaches can provide very different insights on the general behaviour of the numerical scheme. The former is more suitable for temporally evolving problems, whereas the latter is more appropriate for spatially developing problems.

Although such techniques are able to reveal useful numerical properties, they have been often categorised as too simplistic. In fact, a direct connection between the numerical discretisation of the linear advection equation and more complex three-dimensional nonlinear simulations can be everything but trivial.
The high-order schemes' research community has also produced substantial works on generalised forms of spectral analyses.
This includes analyses of linear advection on multiple dimensions~\cite{lele1992compact,trojak2020effect,saez2021spectral}, of linear advection problems with non-constant velocity~\cite{manzanero2018dispersion,manzanero2017insights,tonicello2021comparative} and even of alternative eigenanalysis techniques, such as non-modal analysis~\cite{Fernandez_2019} and combined-mode analysis~\cite{alhawwary2020combined}.
Nonetheless, all of these approaches commonly involve a semi-discrete form of the linear advection equation, where only spatial numerical errors are considered. The main goal of the present work is focused on a fully-discrete formulation of spatial eigenanalysis where also temporal errors are taken into account.
Despite the fact that some recent work has been presented regarding fully-discrete temporal eigenanalysis~\cite{yang2013dispersion,vermeire2017behaviour,alhawwary2018fourier}, the full generalisation of spatial eigenanalysis including a temporal discretisation remains, as of yet, unexplored.
Even if spatial errors play a critical role in the simulation of turbulent flows, the interaction with the temporal counterparts is everything but obvious. The temporal discretisation can significantly affect the spatial errors, leading to much different results in terms of accuracy and stability of the numerical simulation.

The ultimate goal of simplified spectral analyses is to predict, within a reasonable range, how a certain scheme, for a given resolution, would behave in practical flow configurations. For instance, it is possible to estimate which grid resolution allows one to reliably perform Implicit LES (ILES), thereby reducing the computational burden otherwise incurred. In the same way, temporal discretisations would benefit from large time steps, while introducing negligible numerical errors. In other words, a trade-off between accuracy and computational cost needs to be found both spatially and temporally. Finding such delicate balance using fully three-dimensional simulations can be extremely expensive and even pointless. Ideally, efficient and reliable simplified spectral analysis techniques aim at providing precisely this kind of information at very low computational cost.

The paper is organised as follows. In section~\ref{S:2}, we introduce the fully-discrete eigenanalysis framework which serves both temporal and spatial approaches. This framework is applied to the Flux Reconstruction method, which can recover both the discontinuous Galerkin and Spectral difference schemes. In section~\ref{S:4}, we present the eigenanalysis results for the fully-discrete spatial approach, where we focus on the interpretation of the diffusion curves and the contributions of the different (transmitted/reflected) modes. The observations will be specifically classified into well-resolved and under-resolved frequency ranges, providing a direct link with DNS and LES approaches, respectively. In section~\ref{S:5}, we present the numerical experiments conducted to assess the validity of the eigenanalysis. In particular, first a discretisation of the linear advection equation will be considered, followed by more complex inviscid flows based on the Euler equations. In line with the theoretical discussion, numerical results will be arranged in well-resolved and under-resolved flows. Finally, in section~\ref{S:6}, we outline the key conclusions of this work.

\section{Eigenanalysis framework}\label{S:2}
In this section, we introduce the fully-discrete eigenanalysis framework, which serves both the temporal (section~\ref{subsec:temporal}) and the spatial (section~\ref{subsec:spatial}) approaches. The classical fully-discrete temporal eigenanalysis (that was originally reported by Vermeire and Vincent~\cite{vermeire2017behaviour}) is presented first, as a way of introducing the key building blocks for the fully-discrete spatial eigenanalysis.

\subsection{Classical fully-discrete temporal eigenanalysis}\label{subsec:temporal}
All the relevant steps of temporal eigenanalysis will be herein introduced using the FR scheme, which is well-known for its ability to recover different methods with a specific choice of its correction functions~\cite{de2014connections,mengaldo2015discontinuous,mengaldo2016connections}. The present study considers two such methods in particular, namely DG and SD. In fact, this strategy allows one to perform subsequent eigenanalysis studies for any other method that FR is capable of recovering.

In the standard temporal analysis, the linear advection equation is discretised looking for wave-like solutions in order to study the numerical description of specific wavenumbers and their evolution over time. 
Then, dispersion and dissipation properties of any scheme follow directly from the corresponding eigensolutions. 
Let us now suppose the case of a constant unitary advection velocity: 
\begin{equation}
\pd{u}{t}+\pd{u}{x}=0.
\label{advection}
\end{equation}
This equation admits plane wave solutions of the form 
\begin{equation}
u(x,t)=e^{\iota(\theta x-\omega t)},\quad \text{with $\iota = \sqrt{-1}$},
\label{wave}
\end{equation}
provided that the angular frequency $\omega=\omega(\theta)$ is such that
\begin{equation}
\textrm{Re}(\omega)=\theta \qquad \textrm{and} \qquad \textrm{Im}(\omega)=0, 
\label{ReIm}
\end{equation}
where $\theta$ is a real-valued wavenumber chosen at the initial condition (in the case of standard temporal eigenanalysis).
Equation~\eqref{ReIm} provides what are respectively known as dispersion and diffusion relations for the exact plane wave solution.
A numerical discretisation of the linear advection equation is now introduced on a grid of equally spaced elements of fixed width $h_{n}=h=1$.
Under such assumptions, a FR spatial discretisation of equation~\eqref{advection} within the standard element $\Omega_{n}$ can be written as
\begin{equation}
\der{\hat{u}_{n}}{\textrm{t}} = -2 \sum_{j=0}^{N} \hat{u}_{n}\der{l_{j}}{\hat{x}}(\hat{x}_{i})-(2\hat{f}^{I}_{L}-2 \hat{u}_{L})\der{g_{L}}{\hat{x}}(\hat{x}_{i})-(2\hat{f}^{I}_{R}-2\hat{u}_{R})\der{g_{R}}{\hat{x}}(\hat{x}_{i}),
\label{discre}
\end{equation}
where $N$ is the order of the solution polynomial, $\hat{u}_{L}=\hat{u}_{n}(-1,t)$ and $\hat{u}_{R}=\hat{u}_{n}(+1,t)$. 

To examine a sufficiently general framework, the classical set of numerical fluxes for the linear advection equation is considered: 
\begin{equation}
\hat{f}_{L}^{I}=(1-\alpha)\hat{u}_{n-1}(+1,t)+\alpha \hat{u}_{n}(-1,t)
\label{eq:l:flux}
\end{equation}
and
\begin{equation}
\hat{f}_{R}^{I}=(1-\alpha)\hat{u}_{n}(+1,t)+\alpha \hat{u}_{n+1}(-1,t).
\label{eq:r:flux}
\end{equation}
Note that $\alpha=0$ leads to a standard upwind scheme, whereas $\alpha=0.5$ leads to a centred scheme. In the following sections, unless stated otherwise, standard upwind fluxes will be considered.
The $\alpha$ parameter can be used to mimic different levels of interface upwinding, which can be directly linked to more complex numerical fluxes for the Euler or Navier-Stokes equations (via Riemann solvers). 
Equation~\eqref{discre} can be written in matrix form as:
\begin{equation}
\der{\hat{\textbf{u}}^{n}}{\textrm{t}} = -2 \textbf{D}\hat{\textbf{u}}^{n}-(2\hat{f}^{I}_{L}-2 \textbf{l}^{T} \hat{\textbf{u}}^{n})\textbf{g}_{L}-(2\hat{f}^{I}_{R}-2 \textbf{r}^{T} \hat{\textbf{u}}^{n})\textbf{g}_{R},
\end{equation}
where $\hat{\textbf{u}}^{n}_{i}=\hat{u}(\hat{x}_{i})$, $\textbf{D}_{ij}=\der{l_{j}}{\hat{x}}(\hat{x}_{i})$, $\textbf{g}^{L/R}_{i}=\der{g_{L/R}}{\hat{x}}(\hat{x}_{i})$ and $\textbf{r}_{i}=l_{i}(1)$, $\textbf{l}_{i}=l_{i}(-1)$. 
 
In order to reduce the infinite dimensional problem to an element-wise framework, Bloch wave-like type of solutions are sought, namely:
\begin{equation}
\hat{\textbf{u}}^{n}=e^{\iota(\tilde{\theta}x_{n}/h-\tilde{\omega}t)}\hat{\textbf{v}},
\label{bloch}
\end{equation}
where tilde accent denotes the numerical counterparts of continuum variables and $x_{n}$ indicates the left boundary of the element $n$ in the global reference frame. 
Once $\hat{\textbf{u}}^{n}$ is properly defined, due to periodicity, it is straightforward to obtain a closed form for all the fluxes expressed in equation~\eqref{discre}: 
\begin{equation}
\begin{aligned}
\hat{f}_{L}^{I}&=(1-\alpha)e^{-\iota \tilde{\theta}}&\textbf{r}^{T}\hat{\textbf{u}}^{n}&+\alpha \textbf{l}^{T}&\hat{\textbf{u}}^{n}, \\
\hat{f}_{R}^{I}&=(1-\alpha)&\textbf{r}^{T}\hat{\textbf{u}}^{n}&+\alpha \textbf{l}^{T}e^{\iota \tilde{\theta}}&\hat{\textbf{u}}^{n}. \\
\end{aligned}
\end{equation}
Finally, it can be seen that, due to solution periodicity, the problem has now been reduced to an element-wise formulation, \ie, the numerical fluxes now depend only on the local solution $\hat{\textbf{u}}^{n}$.
Hence, index $n$ can be dropped and the final discretised equation can be written as
\[
\der{\hat{\textbf{u}}}{\textrm{t}}= -2\bigg(\textbf{D}\hat{\textbf{u}}+((1-\alpha)e^{-\iota \tilde{\theta}}\textbf{r}^{T}\hat{\textbf{u}}+\alpha \textbf{l}^{T}\hat{\textbf{u}}-\textbf{l}^{T} \hat{\textbf{u}})\textbf{g}_{L}+((1-\alpha)\textbf{r}^{T}\hat{\textbf{u}}+\alpha \textbf{l}^{T}e^{\iota \tilde{\theta}}\hat{\textbf{u}}-\textbf{r}^{T} \hat{\textbf{u}})\textbf{g}_{R}\bigg),
\]
or, in a more compact way:
\begin{equation}
\der{\hat{\textbf{u}}}{\textrm{t}} = -2(\textbf{C}^{-}e^{-\iota \tilde{\theta}}+\textbf{C}^{0} +\textbf{C}^{+}e^{\iota \tilde{\theta}})\hat{\textbf{u}} ,
\label{final}
\end{equation}
where $\textbf{C}^{0}=\textbf{D}-(\alpha-1)\textbf{g}_{L}\textbf{l}^{T}-\alpha \textbf{g}_{R}\textbf{r}^{T}$, $\textbf{C}^{-}=(1-\alpha)\textbf{g}_{L}\textbf{r}^{T}$ and $\textbf{C}^{+}=\alpha \textbf{g}_{R}\textbf{l}^{T}$. When FR is made to recover specific methods (such as DG or SD), matrices $\textbf{C}$ above are what will differentiate them, but the structure of the problem does not change.
%
%
%
%
%
%
%
In either case, the discretised system is now reduced to the following linear dynamical system:
\begin{equation}
\der{\hat{\textbf{u}}}{\textrm{t}}= \textbf{H}(\tilde{\theta})\hat{\textbf{u}}.
\label{eigen}
\end{equation}
%

The approach herein outlined represents a classical tool in numerical analysis, particularly used for the theoretical study of novel high-order schemes. Notice that the present analysis is often called \emph{semi-discrete} temporal eigenanalysis due to the absence of time discretisation. Indeed, taking advantage of the analytical form of wave-like solutions, the dynamical system defined by equation~\eqref{eigen} can be reduced to the following eigenvalue problem:
\begin{equation}
-\iota \tilde{\omega}  \hat{\textbf{u}}= \textbf{H}(\tilde{\theta}) \hat{\textbf{u}}.
\label{eq:semi-discrete_temp}
\end{equation}
Alternatively, instead of considering the analytical form of the wave-like solution, a numerical time integration scheme can be introduced to advance in time. The numerical solution at the time step $t+1$ can be generally approximated by the M-stage explicit Runge-Kutta schemes as
\begin{equation}
\hat{\textbf{u}}^{t+1}  \approx \sum_{m=0}^{M}\beta_{m}\frac{(\textbf{H}(\tilde{\theta}) \Delta t)^{m}}{m!} \hat{\textbf{u}}^{t} = \textbf{T}(\tilde{\theta},\Delta t)\hat{\textbf{u}}^{t},
\label{eq:RK_adv}
\end{equation}
where $\Delta t$ is the time step and $\boldsymbol{\beta}$ contains the coefficients of the specific Runge-Kutta scheme.

This formulation is denoted as \emph{fully-discrete} since both spatial and temporal discretisations are simultaneously involved. Considering the present generalised formulation, two main closely linked analyses can be performed. The first is relatively popular and originates from the classical stability analysis of numerical discretisation of partial differential equations. In particular, it is well-known that in equation~\eqref{eq:RK_adv}, the combined space-time discretisation is denoted as \emph{stable} if the spectral radius of the iteration matrix $\textbf{T}$ is strictly smaller than unity. Namely, for the specific case of Von Neumann analysis, the full discretisation is considered linearly stable if, for a given time step $\Delta t$, 
\begin{equation}
\forall \, \tilde{\theta} \in \big[-\pi (N+1)/h; \quad  \pi (N+1)/h \big]: \quad \lambda(\textbf{T}(\tilde{\theta},\Delta t)) < 1,
\label{eq:stab}
\end{equation}
where $\lambda(\textbf{T})$ denotes the spectral radius of the matrix $\textbf{T}$. 

A complementary analysis, instead, takes into account the Bloch wave-like form of the numerical solution and generalises the concept of dispersion/diffusion curves in the fully-discrete analyses. In fact, exploiting the Bloch-like form of the approximate solution, it can be shown (see~\cite{asthana2015high,vermeire2017behaviour}) that
\begin{equation}
\hat{\textbf{u}}^{t+1}  = e^{-\iota \tilde{\omega} \Delta t} \hat{\textbf{u}}^{t}.
\end{equation}
Consequently, the fully-discrete version of the dispersion/diffusion eigenanalysis can be written as:
\begin{equation}
e^{-\iota \tilde{\omega}  \Delta t} \hat{\textbf{u}}^{t} = \textbf{T}(\tilde{\theta},\Delta t)\hat{\textbf{u}}^{t}. 
\label{eq:fully-discrete_temp}
\end{equation}
Notice, in fact, that for $\Delta t \to 0$ we have
\begin{equation*}
\begin{aligned}
\textbf{T} &\rightarrow & \textbf{I} + \textbf{H} \Delta t + O(\Delta t^{2}), \\ 
e^{-\iota\tilde{\omega}  \Delta t} & \rightarrow & 1 - \iota \omega \Delta t + O(\Delta t^{2}) \\
\end{aligned}
\end{equation*}
and the problem reduces to 
\begin{equation*} 
\begin{aligned}
(1- \iota \tilde{\omega}  \Delta t + O(\Delta t^{2}))\hat{\textbf{u}}^{t}&=  (\textbf{I} + \textbf{H} \Delta t + O(\Delta t^{2}))\hat{\textbf{u}}^{t} \\ 
-\iota \tilde{\omega}  \Delta t \hat{\textbf{u}}^{t}&= \textbf{H} \Delta t \hat{\textbf{u}}^{t} + O(\Delta t^{2}) \\
-\iota \tilde{\omega}  \hat{\textbf{u}}^{t}&= \textbf{H} \hat{\textbf{u}}^{t} \\
\end{aligned}
\end{equation*}
which coincides with classical semi-discrete eigenvalue problem. 

In a similar way with respect to the classical analysis, considering \eqref{eq:fully-discrete_temp}, it is possible to compute the frequencies $\tilde{\omega}$ for any given value of $\tilde{\theta}$ and $\Delta t$. Therefore, the choice of the time integration scheme and, in particular, of a stable time step $\Delta t$, will affect the dispersion and diffusion curves. Such curves can be interpreted in the same manner as in the semi-discrete analysis and significant deviations from the semi-discrete dispersion/diffusion curves can crucially affect the numerical properties of the spatial discretisation.

We finally note that, had we chosen to work in the equations above with unspecified values of the advection velocity $a$ and the mesh spacing $h$, the relevant time-step factor appearing in equation \eqref{eq:RK_adv} would have been $\overline{\Delta t} = a \Delta t / h$---namely, the Courant–Friedrichs–Lewy (CFL) number---instead of simply $\Delta t$. Hence, we remark that, although we chose to work with $\Delta t$  for the sake of simplicity, one can actually have  in mind the CFL number whenever $\Delta t$ values are mentioned in the remainder of the paper.

\subsection{Fully-discrete spatial eigenanalysis}\label{subsec:spatial}
In the classical temporal eigenanalysis, for any real-valued wavenumber $\tilde{\theta}$, the numerical angular frequency $\tilde{\omega}$ can be obtained through the eigenvalue problem in equation~\eqref{eq:fully-discrete_temp}. The frequency $\tilde{\omega}$ will describe the numerical evolution of the specified wavenumber $\tilde{\theta}$ over time. It is however possible to invert the problem by imposing a time frequency $\tilde{\omega}$ and subsequently seek for wavenumbers $\tilde{\theta}$ that solve the inverted problem. Such approach is commonly referred to as \emph{spatial eigenanalysis} and it represents a very useful tool for non-periodic problems since it aims at studying the dynamics of prescribed time frequencies over space.

Within the framework of semi-discrete eigenanalysis, the equation defining the relationship between $\tilde{\theta}$ and $\tilde{\omega}$ is the same, namely equation~\eqref{eq:semi-discrete_temp}, and  can be written in an alternative form as
\begin{equation}
\det(\textbf{H}(\tilde{\theta}) + \iota \tilde{\omega}\textbf{I}) = 0.
\label{eq:semi-discrete_spat}
\end{equation}
In this case, $\tilde{\omega}$ is a real-valued prescribed scalar and the nonlinear equation~\eqref{eq:semi-discrete_spat} is solved in the unknown $\tilde{\theta}$. 
This nonlinear equation can be particularly complex to solve and, except for some very particular cases, cannot be analytically inverted.
Considering the fully-discrete space-time discretisation, an equivalent equation which defines the relation between $\tilde{\omega}$ and $\tilde{\theta}$ can be derived from equation~\eqref{eq:fully-discrete_temp}, namely
\begin{equation}
\det(\textbf{T}(\tilde{\theta},\Delta t )-e^{-\iota \tilde{\omega} \Delta t} \textbf{I})=0.
\label{eq:fully-discrete_spat}
\end{equation}

Note of course that, in the temporal eigenanalysis, the addition of a temporal discretisation does not significantly complicate the overall problem. In fact, the matrix $\textbf{T}$ can be easily computed once $\tilde{\theta}$ and $\Delta t$ are prescribed. Subsequently, the eigenvalue problem can be solved in $e^{-\iota \tilde{\omega} \Delta t}$, which can be finally inverted to obtain the exact value of $\tilde{\omega}$. However, in the spatial eigenanalysis, the specific matrix $\textbf{T}$ influences the general structure of the nonlinear equation to be solved in $\tilde{\theta}$. In particular, if the simple explicit Euler integration scheme is considered, then equation~\eqref{eq:fully-discrete_spat} becomes:
\begin{equation}
\det(\textbf{I} + \Delta t \textbf{H}(\tilde{\theta},\Delta t) - e^{-\iota \tilde{\omega} \Delta t} \textbf{I})=0.
\end{equation}
If $z$ is defined as $z=e^{\iota \tilde{\theta}}$, the final nonlinear equation in $z$ can be written as 
\begin{equation}
\det((1-e^{-\iota \tilde{\omega} \Delta t})\textbf{I} -2 \Delta t (\textbf{C}^{-}z^{-1}+\textbf{C}^{0} +\textbf{C}^{+}z))=0.
\label{eq:fully-discrete_spat_euler}
\end{equation}

Again, notice that for $\Delta t \to 0$, the diagonal term  $1-e^{-\iota \tilde{\omega} \Delta t}$ tends to $\iota \tilde{\omega} \Delta t$ and the fully-discrete formulation goes back to the classical semi-discrete form:
\begin{equation}
\det(\iota \tilde{\omega}\textbf{I} - 2(\textbf{C}^{-}z^{-1}+\textbf{C}^{0} +\textbf{C}^{+}z))=0,
\label{spatial_eigen}
\end{equation}
which is exactly equation~\eqref{eq:semi-discrete_spat}. 
On the other hand, for a RK22 scheme, the formulation becomes much more complex as higher powers of $z$ appear in the final equation:
\begin{equation}
\begin{aligned}
\det \bigg(&(1-e^{-\iota \tilde{\omega} \Delta t})\textbf{I}+ \\
& + \Delta t^{2} \big(\textbf{C}^{-}\textbf{C}^{-}\big) z^{-2} + \\
& + \big( \Delta t^{2} (\textbf{C}^{-}\textbf{C}^{0} + \textbf{C}^{0}\textbf{C}^{-}) -2 \Delta t \textbf{C}^{-} \big) z^{-1} +\\
& + \big( \Delta t^{2} (\textbf{C}^{-}\textbf{C}^{0}+ \textbf{C}^{0} \textbf{C}^{0} +\textbf{C}^{+}\textbf{C}^{-}) -2 \Delta t \textbf{C}^{0} \big) \\
& + \big( \Delta t^{2} (\textbf{C}^{+}\textbf{C}^{0} + \textbf{C}^{0}\textbf{C}^{+}) -2 \Delta t \textbf{C}^{+} \big) z +\\
 & + \Delta t^{2} \big(\textbf{C}^{+}\textbf{C}^{+}\big) z^{2} \bigg)=0.
\end{aligned}
\label{eq:rootsrk2}
\end{equation}
This way, the order of the nonlinear equation increases significantly and additional roots for $z$ are expected to be found (especially for time discretisation schemes of higher order). The role of multiple wave modes in spatial eigenanalysis has been discussed in more detail in \cite{moura2020spatial}. As properly explained in the next section, only one solution will be denoted as the physical one, whereas the others are considered spurious.

\section{Theoretical Results}\label{S:4}
In this section, we show the results associated with the fully-discrete spatial eigenanalysis just introduced. The theoretical findings of the fully-discrete temporal eigenanalysis previously reported by Vermeire and Vincent~\cite{vermeire2017behaviour} were correctly reproduced within the present implementation too. These have been used for validation purposes only and, for the sake of brevity, are not shown here.

It is important to highlight that, from equation~\eqref{wave},  the numerical wave-like solutions depend on the numerical equivalent of $\theta$, i.e.\ $\tilde{\theta}$ (whose imaginary part makes the diffusion curve plots below) according to
\begin{equation} \label{wave-exp-sign}
    u \propto \exp [ -\mbox{Im} (\tilde{\theta}) x ]  \exp[  \iota  \mbox{Re} (\tilde{\theta}) x - \omega t ] \mbox{ ,}
\end{equation}
where reflected waves should be considered with $\Delta x < 0$, thus being stable for $\mbox{Im} (\tilde{\theta}) < 0$.


In the following, we consider high-order FR-based DG/SD methods in space, coupled with explicit Runge-Kutta schemes of increasing order of accuracy (and increasing number of stages): RK11, RK22, RK33 and RK45. Hence, we start with the $1^\mathrm{st}$-order accurate, one-stage Runge-Kutta scheme (more commonly known as the Explicit Euler method) until finally addressing the well-known $4^\mathrm{th}$-order accurate, five-stages Runge-Kutta scheme by Carpenter and Kennedy \cite{carpenter1994fourth}. This allows for a detailed exposition of the fully-discrete eigenanalysis results, whose complexity grows with the number of stages of the temporal scheme, as shown below.


In fig.~\ref{fig:SD_k4_RK11}, we show the dissipation curve for the $5^{\mathrm{th}}$-order FR-SD standard upwind method when coupled with RK11. This curve is shown for increasing values of the time step $\Delta t$, namely $\Delta t = 0.01, 0.02, \dots, 0.1$, with dissipation values becoming more negative as $\Delta t$ increases. Note that, given a $\Delta t$ value, only one eigenmode/curve exists in this case, corresponding to a single transmitted physical wave. This is because, from eq.~\ref{eq:fully-discrete_spat_euler}, the characteristic polynomial in the unknown $z$ has the same order of the semi-discrete analysis (compare with eq.~\ref{spatial_eigen}). We recall that, in the semi-discrete analysis \cite{mengaldo2018spatial}, only the physical transmitted mode exists in case of standard upwinding ($\alpha = 0$ in eq.~\ref{final}), whereas one reflected spurious wave exists in addition if $\alpha \neq 0$. As for the curves in fig.~\ref{fig:SD_k4_RK11}, negative dissipation values indicate in fact anti-dissipative behaviour (convective instability), since we are dealing with a physical transmitted wave, cf.\ eq.~\ref{wave-exp-sign}. Hence, fig.~\ref{fig:SD_k4_RK11} essentially re-affirms that RK11 (explicit Euler) is only marginally stable for pure advection \cite{lomax2013fundamentals}. Interestingly, for moderately large $\Delta t$, waves regain stability beyond a certain frequency $\omega$, probably due to increased dissipation from the spatial discretisation at large frequencies. These features will be tested in the next section, as they highlight how the fully-discrete eigenanalysis is useful in revealing non-intuitive characteristics stemming from the coupling between spatial and temporal discretisations.


\begin{figure}[t]
\centering
\subfigure{\includegraphics[width=0.46\textwidth]{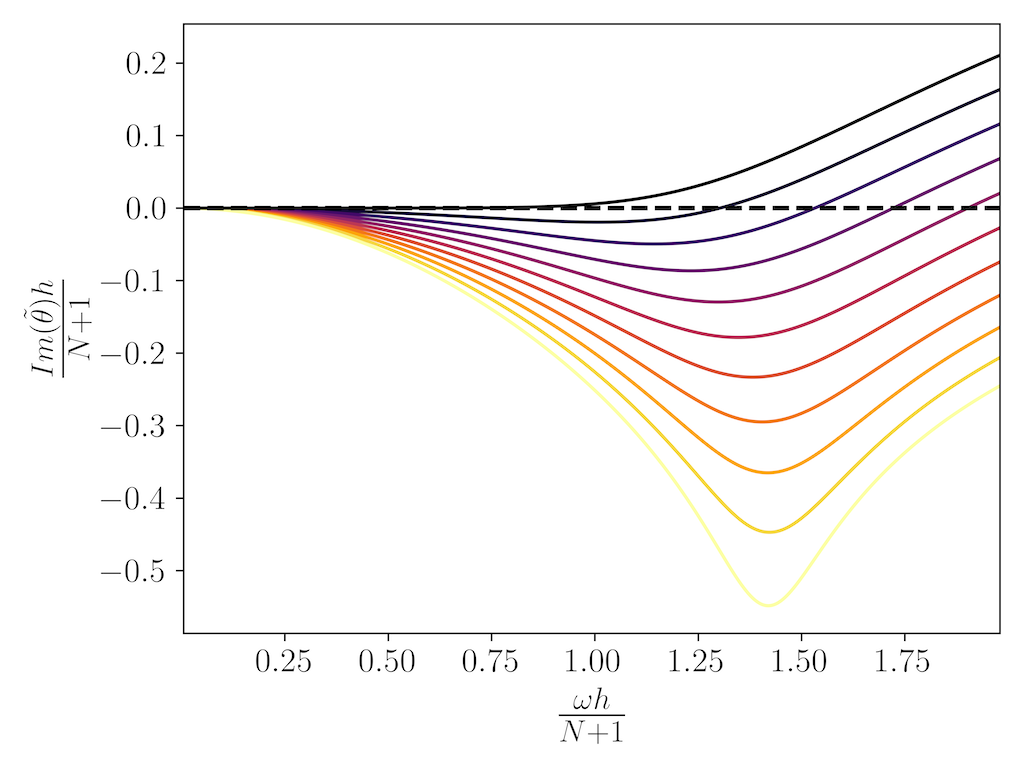}}
\subfigure{\includegraphics[width=0.46\textwidth]{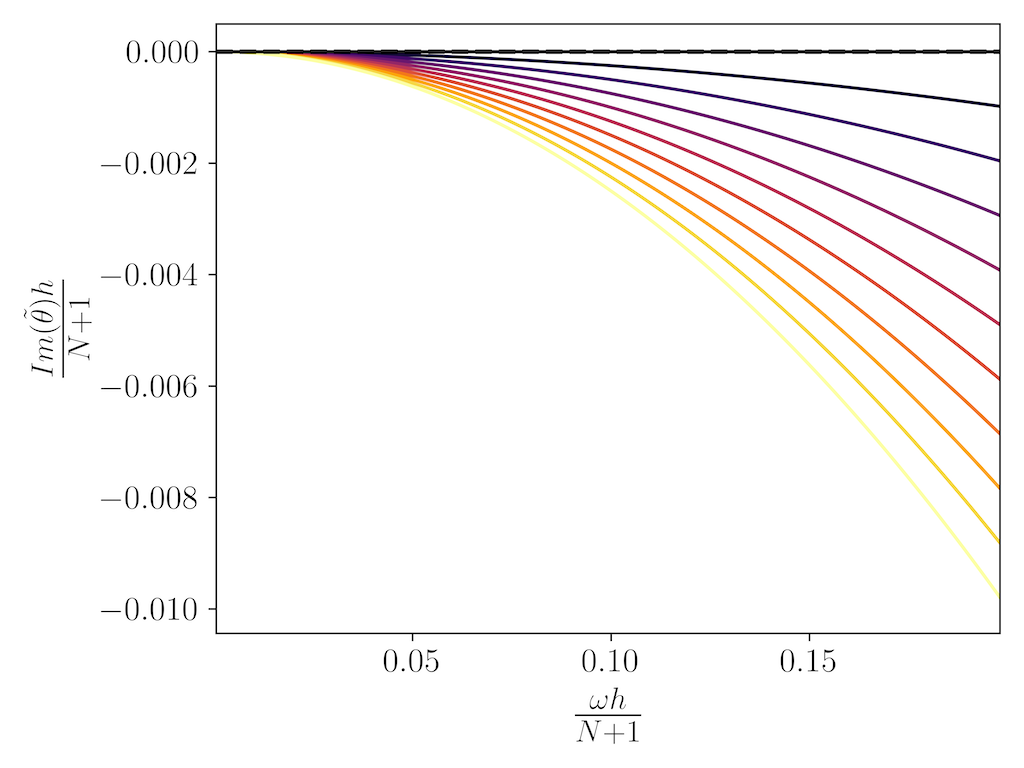}}
\caption{Dissipation curve for the $5^\mathrm{th}$-order FR-SD standard upwind scheme coupled with RK11 (explicit Euler) denoting a single transmitted physical mode, shown for $\Delta t = 0.01, 0.02, \dots, 0.1$ (with dissipation values becoming more negative as $\Delta t$ increases). The plot on the right shows a closer look near the origin. These re-affirm that RK11 is convectively unstable for pure advection.}
\label{fig:SD_k4_RK11}
\end{figure}

\begin{figure}
\centering
\includegraphics[trim=0 0 0 0, width=1.00\textwidth]{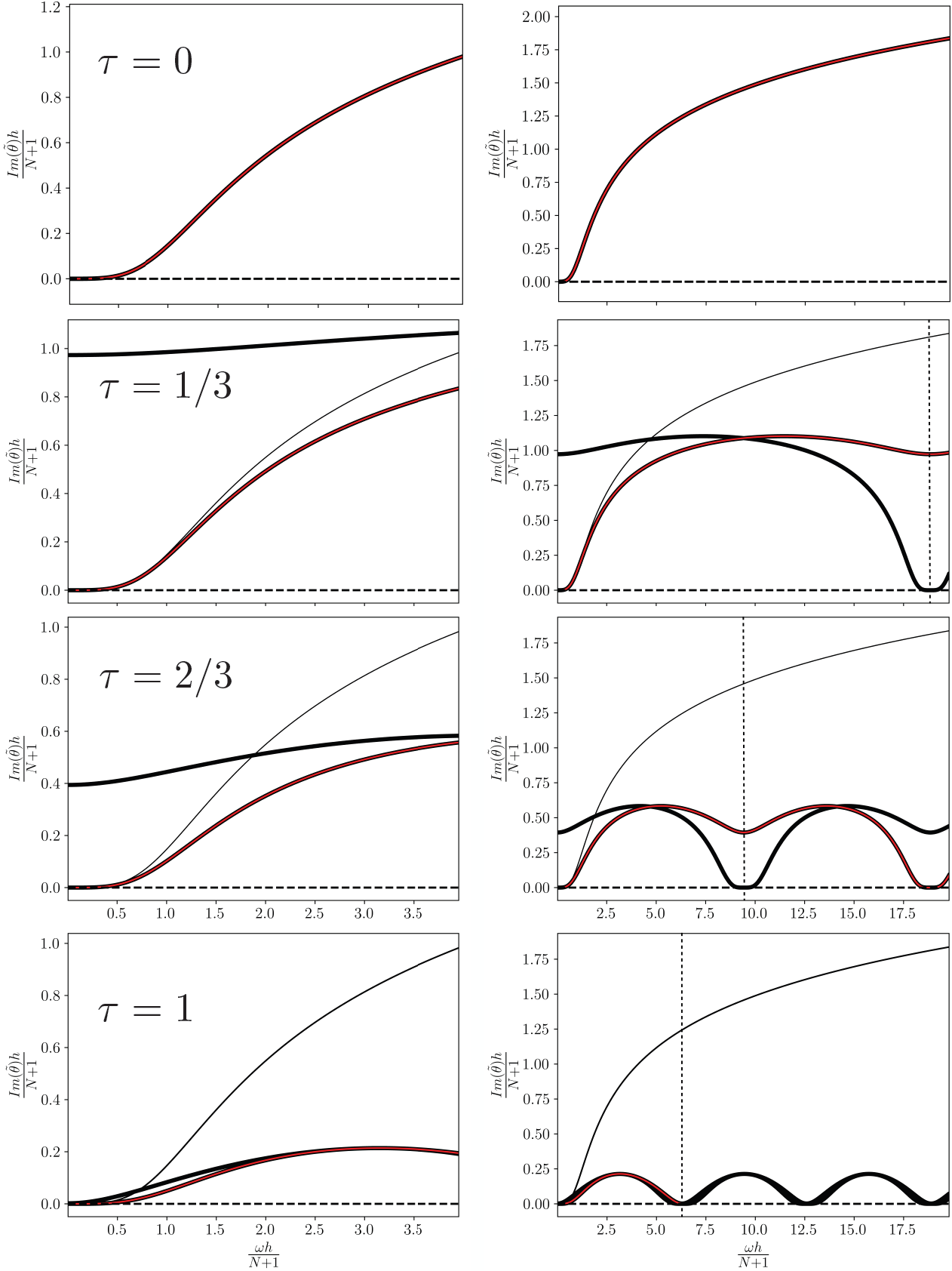}
\caption{Dissipation curves for the $2^\mathrm{nd}$-order FR-SD standard upwind scheme coupled with RK22 for increasing values of $\tau = \Delta t/\Delta t_{\max}$. The plots on the right show the same left curves, but for a larger frequency range. The semi-discrete ($\tau = 0$) result also appears when $\tau > 0$ for reference as a thin continuous curve. The physical mode is shown in red, whereas the spurious transmitted mode is shown as a thick black curve. A vertical dashed line indicates the cut-off frequency $\omega^{c} = \pi / \Delta t$.}
\label{fig:SD_k1_RK22}
\end{figure}

Now we consider the classical $2^\mathrm{nd}$-order explicit Runge-Kutta (RK22) scheme, this time coupled for simplicity with the $2^\mathrm{nd}$-order FR-SD standard upwind method. The resulting discretisation is stable for $\Delta t < \Delta t_{\max}$, where $\Delta t_{\max}$ is the critical time-step size at which instabilities first appear. In fig.~\ref{fig:SD_k1_RK22}, we show the corresponding dissipation curves for increasing values of $\tau = \Delta t/\Delta t_{\max}$. When $\tau=0$, the results match those of the semi-discrete analysis, where only the transmitted physical mode exists in case of standard upwinding. For $\tau > 0$, a spurious transmitted mode appears in addition to the transmitted physical one, not to be confused with the thin continuous curve which simply shows the semi-discrete ($\tau=0$) physical curve as reference (note that the physical curve is shown in red, whereas the spurious one is shown as a thick black curve). The existence of exactly two modes is guaranteed by the characteristic polynomial in eq.~\ref{eq:rootsrk2} being only quadratic in $z$, since matrix $\textbf{C}^{+}$ becomes zero in case of standard upwinding ($\alpha = 0$ in eq.~\ref{final}). This feature can be traced algebraically to the number of stages of the time discretisation adopted. In general, the number of modes (in case of standard upwinding) matches the number of stages of the chosen explicit Runge-Kutta scheme. The physical mode can be identified as the one \emph{anchored} to the origin of the dissipation plots, thus featuring vanishing dissipation levels for well-resolved waves (as expected in simulations of pure advection).


Returning to fig.~\ref{fig:SD_k1_RK22}, we note that, as $\tau$ increases, the physical mode becomes less dissipative overall (at least for $\omega h (N+1)^{-1} > 0.5$, say) until finally touching the horizontal axis, whose crossing (for $\tau > 1$) means anti-dissipation (convective instability). The spurious mode, despite being \emph{infinitely dissipated} for $\tau \rightarrow 0$, also has its dissipation reduced as $\tau$ increases; in fact, for moderately large $\tau$, its dissipation levels are comparable to those of the physical mode. Small dissipation is typically undesirable for spurious modes, since it allows them to survive longer as artificial oscillations in a simulation, potentially affecting not only solution quality, but also numerical stability \cite{mengaldo2018spatial2}. Curiously, as $\tau \rightarrow 1$, the spurious mode's dissipation gradually approaches that of the physical mode, almost matching it at $\tau = 1$. This behaviour closely resembles that of reflected spurious modes which exist for either very low or very high upwinding levels (i.e.\ either centred fluxes or \emph{hyper-upwinding}), cf.\ \cite{mengaldo2018spatial}, except that reflected modes have dissipation values of opposite sign when compared to those of their corresponding physical modes. In any case, for the present scheme, both physical and spurious modes touch the horizontal axis of zero dissipation when $\tau = 1$, becoming therefore unstable for $\Delta t > \Delta t_{\max}$.


Still in fig.~\ref{fig:SD_k1_RK22}, note that the spurious mode's dissipation curve (e.g.\ when considered until $\omega^c = \pi / \Delta t$, such that $\omega^c h (N+1)^{-1}$ is marked as a vertical dashed line) resembles a mirror image of the physical mode's curve. What is more generally true, though, is that the spurious curve is simply a shifted replication of the physical curve, while both are periodic functions of period $2\omega^c$. Interestingly, the periodicity of eigencurves in spatial analysis has not been noted in the literature so far. This is because $\omega^c \rightarrow \infty$ when $\tau \rightarrow 0$, cf.\ fig.~\ref{fig:SD_k1_RK22}, consistent with the fact that the spurious mode becomes \emph{infinitely dissipated} for $\tau \rightarrow 0$, as already mentioned. Therefore, periodicity is only visible in fully-discrete spatial analysis. This is in contrast with temporal analysis, where one sees periodicity already in the semi-discrete case, as well as the shifted replication property amongst multiple eigencurves \cite{moura2015linear,moura2016eigensolution}.

%
%

Although the cut-off value $\omega^c = \pi / \Delta t$ can be rightfully recognized as the Nyquist frequency imposed by the discrete time-stepping even outside of the fully-discrete spatial framework \cite{vermeire2017behaviour}, it is only in the fully-discrete spatial analysis that $\omega^c$ exhibits visual correlation with the periodicity of the eigencurves, as already noticed in fig.~\ref{fig:SD_k1_RK22}. In fact, the normalized version of $\omega^c$, namely $\omega^c h (N+1)^{-1}$, is the counterpart of the wavenumber limit $\theta h (N+1)^{-1} = \pi$ typically used in temporal analysis. Hence, it seems appropriate to plot spatial analysis curves only until $\omega^c$. Nevertheless, as it will become clear in subsequent plots, it so happens that the actual limit observed for explicit Runge-Kutta schemes of $R$ stages is $R \, \omega^c / 2$. Interestingly, the conclusion is that multiple stages seem to affect the temporal resolution as if multiple single-stage time steps of size $\delta t =  2 \Delta t / R$ had been employed instead. This seems akin to how multiple element-wise polynomial modes increase the spatial resolution of a single element, as if it had \emph{sub-elements}.

\begin{figure}[t]
\centering
\includegraphics[width=0.6\textwidth]{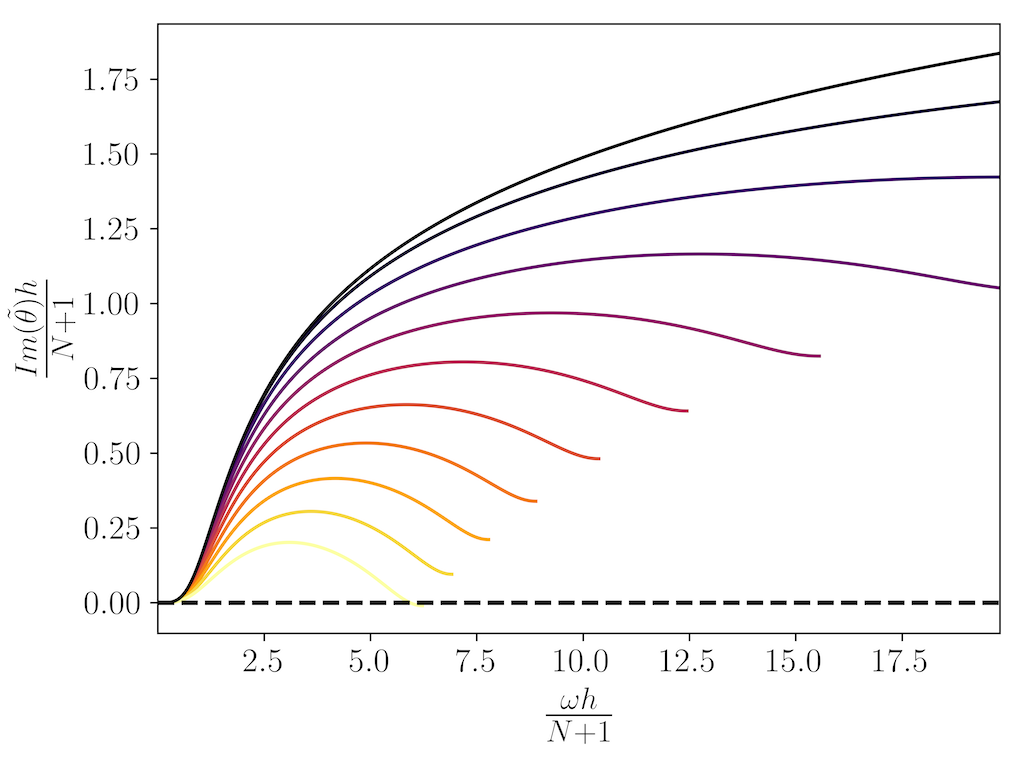}
\caption{Dissipation curves of the physical transmitted mode at increasing values of $\tau$ (from $\tau = 0$ in black until $\tau = 1$ in yellow). Results obtained for the $2^{\mathrm{nd}}$-order FR-SD standard upwind discretisation coupled with RK22.}
\label{fig:SD_k1_RK22_dt}
\end{figure}

A first obvious consequence of adopting $R \, \omega^c / 2$ as the right-most value of $\omega$ when plotting eigencurves is that this bound tends to infinity for $\tau \rightarrow 0$. Indeed, in the semi-discrete spatial analysis, every value of $\omega$ is allowed, in principle. Secondly, such bound gets closer and closer to zero for larger time steps, as only lower and lower frequencies can be solved by the temporal discretisation when the time step increases. These effects are highlighted in fig.~\ref{fig:SD_k1_RK22_dt}, where only the physical mode is represented for different values of $\tau$.

\begin{figure}[t]
\centering
\subfigure{
\includegraphics[width=0.46\textwidth]{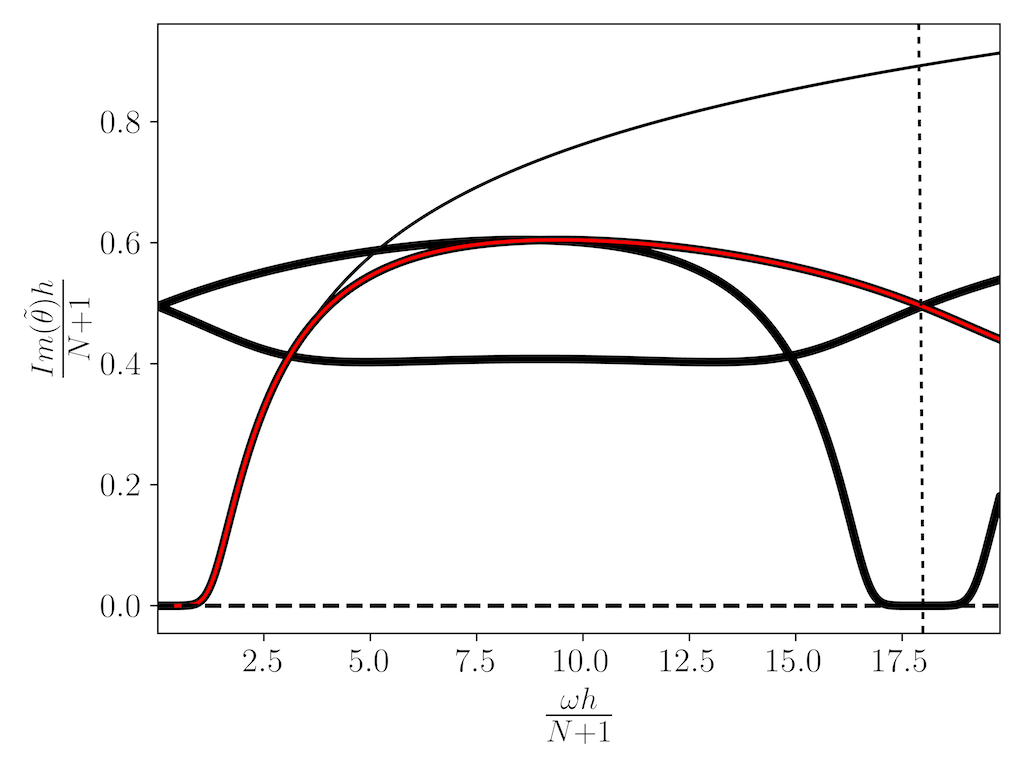}}
\subfigure{
\includegraphics[width=0.46\textwidth]{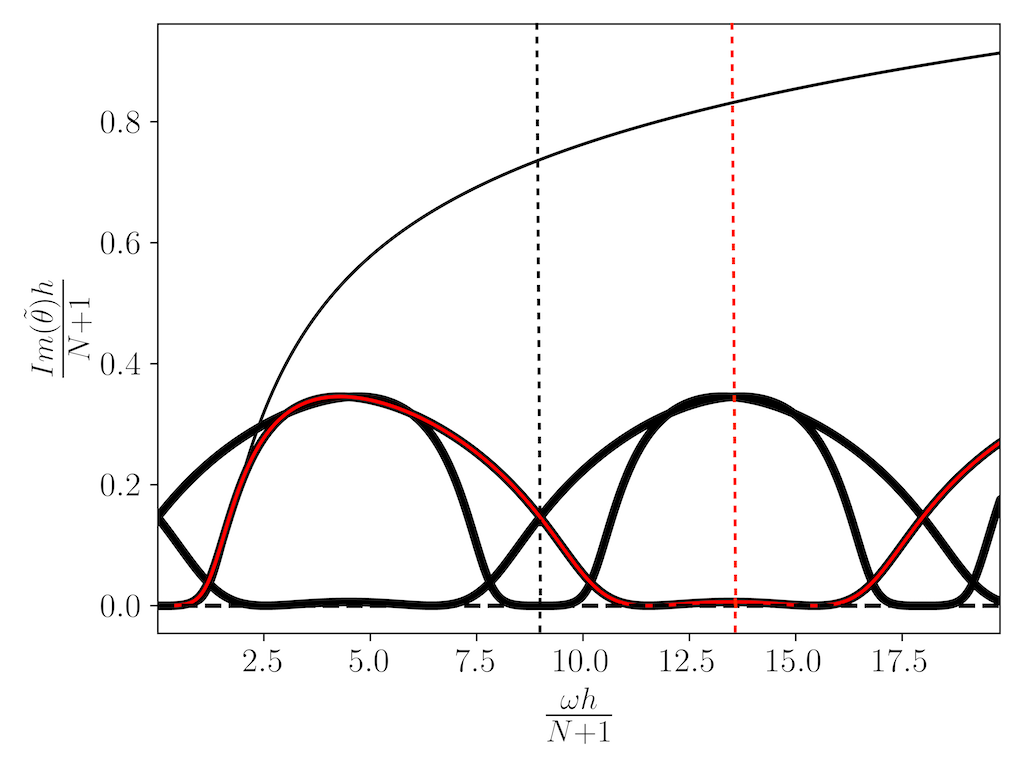} \label{SD_k4_RK45_dt10}}
\caption{Dissipation curves for the $5^\mathrm{th}$-order standard upwind FR-SD coupled with RK33 at $\tau = 0.5$ (left) and $\tau = 1$ (right). The physical mode is shown as a continuous red curve. The vertical black dashed line marks $\omega^{c}$, whereas the vertical red dashed line marks $R \, \omega^{c}/2$ (indicating half a period).}
\label{SD_k4_RK33_dt}
\end{figure}

\begin{figure}[t]
\centering
\subfigure{
\includegraphics[width=0.46\textwidth]{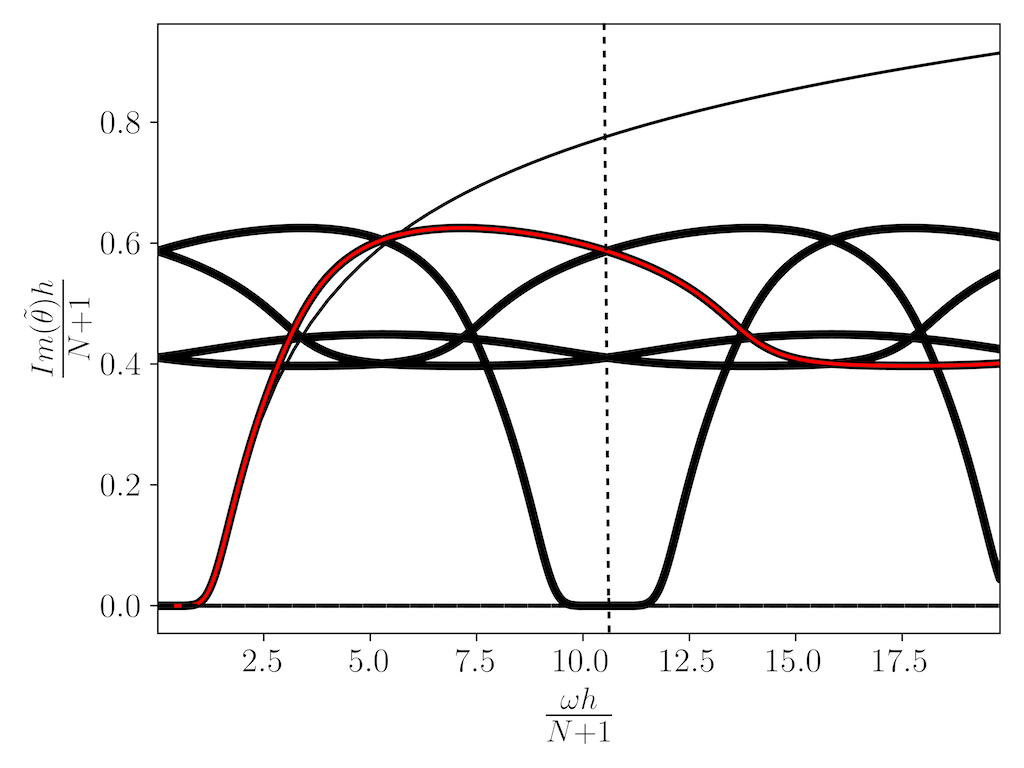}}
\subfigure{
\includegraphics[width=0.46\textwidth]{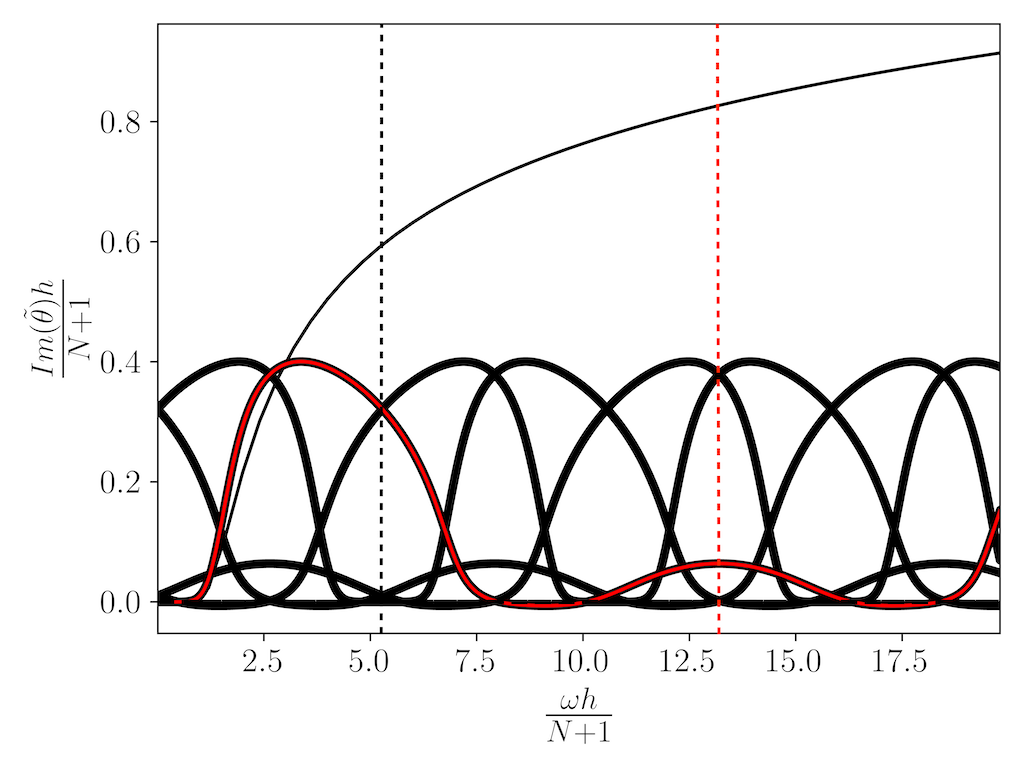} \label{SD_k4_RK45_dt10}}
\caption{Dissipation curves for the $5^\mathrm{th}$-order standard upwind FR-SD coupled with RK45 at $\tau = 0.5$ (left) and $\tau = 1$ (right). The physical mode is shown as a continuous red curve. The vertical black dashed line marks $\omega^{c}$, whereas the vertical red dashed line marks $R \, \omega^{c}/2$ (indicating half a period).}
\label{SD_k4_RK45_dt}
\end{figure}

We now consider the $5^{\mathrm{th}}$-order FR-SD standard upwind scheme coupled with either RK33 (fig.~\ref{SD_k4_RK33_dt}) or RK45 (fig.~\ref{SD_k4_RK45_dt}). Dissipation curves for $\tau = 0.5$ and $\tau = 1$ are respectively shown on the left and right plots within these figures. As previously mentioned, the number of roots in the respective characteristic polynomials matches the number of stages in the adopted time scheme. This is why RK33 (fig.~\ref{SD_k4_RK33_dt}) yields three transmitted modes, whereas RK45 (fig.~\ref{SD_k4_RK45_dt}) yields five. The correspondence between the number of eigenmodes and stages in spatial analysis resembles that between eigenmodes and element-wise polynomial modes in temporal analysis (in line with the previous comment regarding the interpretation of stages as sub-divisions of the actual time step $\Delta t$). In figs.~\ref{SD_k4_RK33_dt} and \ref{SD_k4_RK45_dt}, the black and red vertical dashed lines indicate, respectively, $\omega^c$ and $R \, \omega^c / 2$, the latter matching half the period of the physical mode, as expected. Note that the physical mode is found to be symmetric with respect to the red dashed line (actual Nyquist frequency). Moreover, all eigenmodes are observed to be shifted replications of the physical mode, similar to temporal analysis. This is sometimes useful for the identification of the physical mode.

\begin{figure}[t]
\centering
\subfigure{
\includegraphics[width=0.46\textwidth]{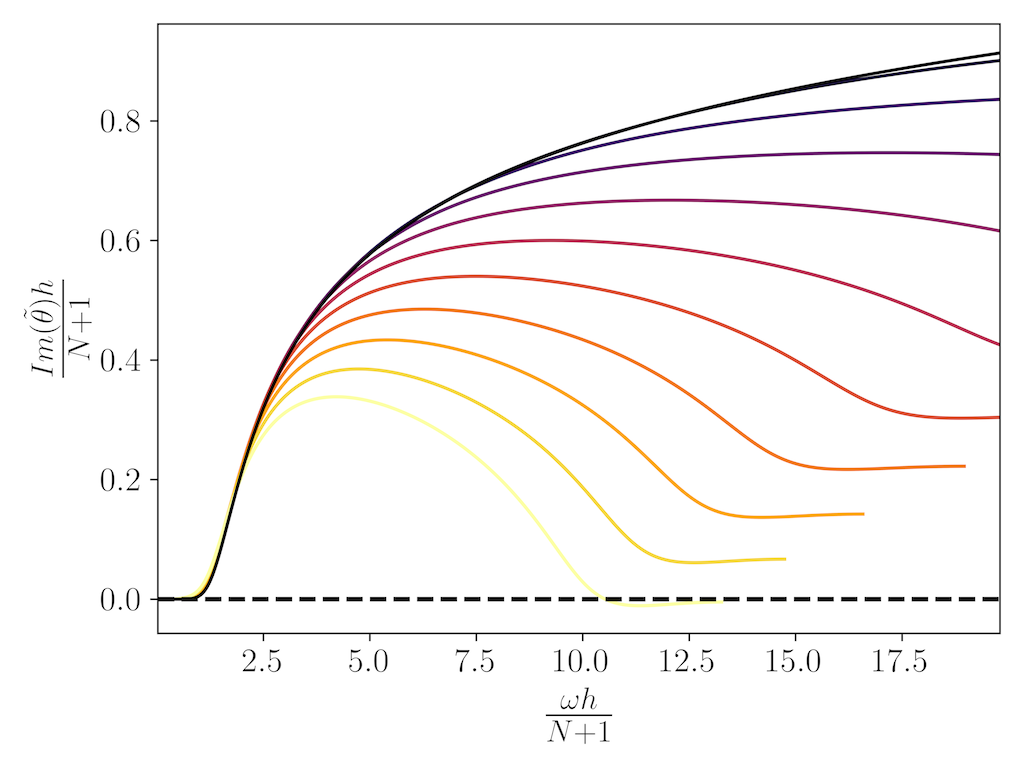}}
\subfigure{
\includegraphics[width=0.46\textwidth]{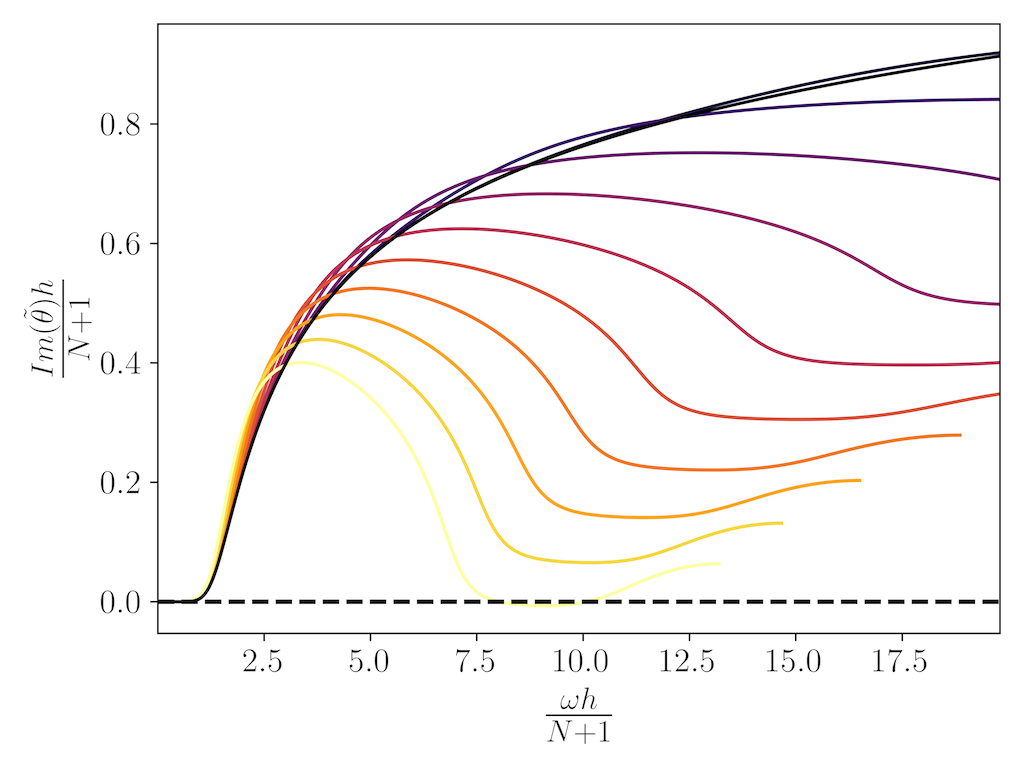}}
\caption{Dissipation curves of the physical transmitted mode at increasing values of $\tau$ (from $\tau = 0$ in black until $\tau = 1$ in yellow). Results obtained for the $5^\mathrm{th}$-order FR-SD standard upwind discretisation coupled with either RK33 (left) or RK45 (right).}
\label{SD_k4_RK}
\end{figure}
\begin{figure}[t]
\centering
\subfigure{
\includegraphics[width=0.46\textwidth]{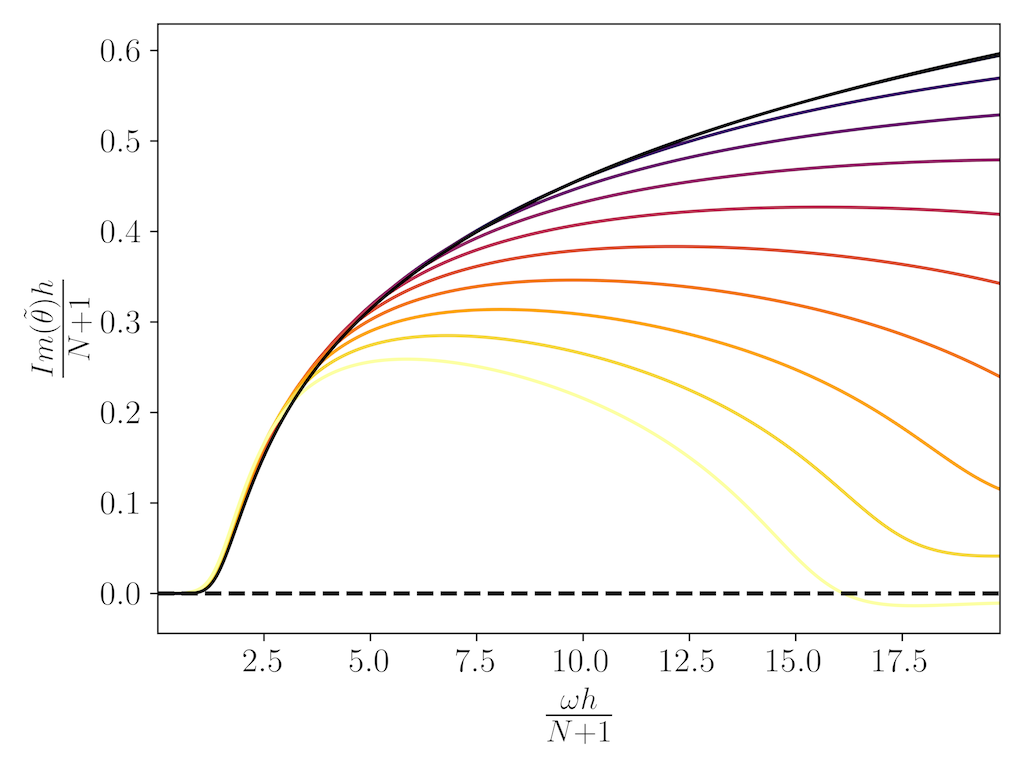}}
\subfigure{
\includegraphics[width=0.46\textwidth]{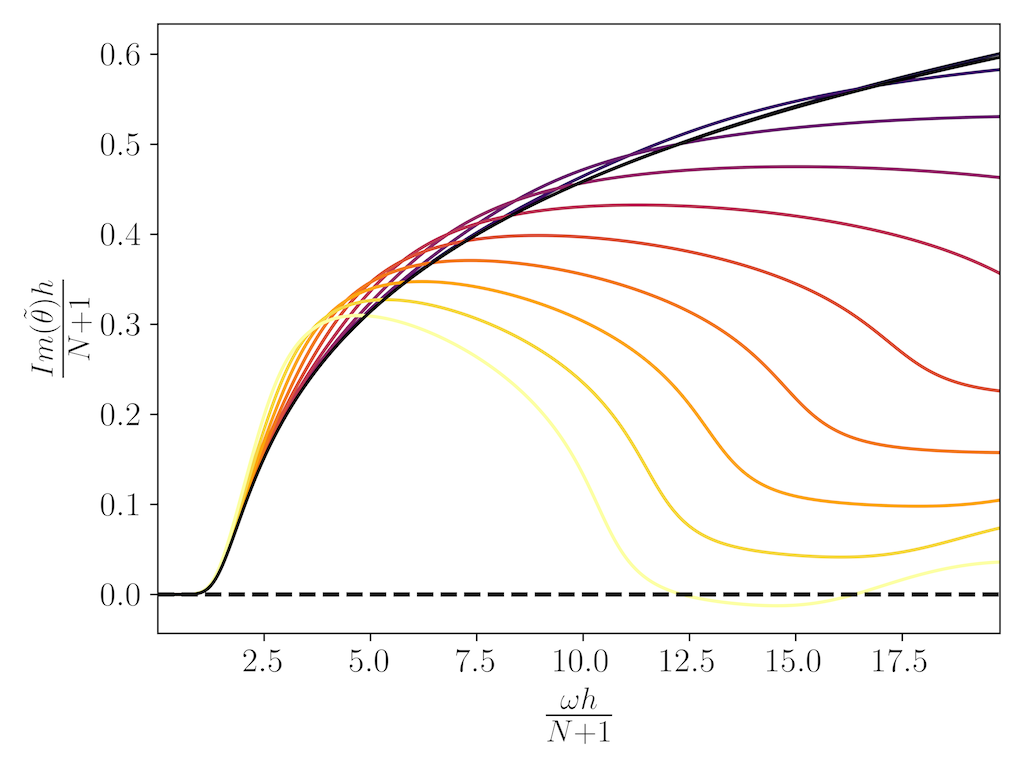}}
\caption{Dissipation curves of the physical transmitted mode at increasing values of $\tau$ (from $\tau = 0$ in black until $\tau = 1$ in yellow). Results obtained for the $5^\mathrm{th}$-order FR-DG standard upwind discretisation coupled with either RK33 (left) or RK45 (right).}
\label{DG_k4_RK}
\end{figure}

The physical curves of figs.~\ref{SD_k4_RK33_dt} and \ref{SD_k4_RK45_dt} have been isolated and are shown in fig.~\ref{SD_k4_RK} for different values of $\tau$. For comparison, the equivalent plots for FR-DG are given in fig.~\ref{DG_k4_RK}. As one can see, the general trends observed for FR-SD are also seen for FR-DG. Some differences, however, can be discerned. For example, since the CFL limit is more restrictive for FR-DG than for FR-SD (both under standard upwinding), the dissipative curves of FR-DG extend to higher frequencies (recall that $\omega^c = \pi / \Delta t$). Moreover, as already pointed out in previous works \cite{hu2002eigensolution,mengaldo2018spatial,mengaldo2018spatial2,tonicello2021comparative}, FR-DG is spatially less dissipative than FR-SD, overall. A more complete comparison between FR-SD and FR-DG is shown in the Appendix (cf.\ figs.~\ref{fig:DGcompile} and \ref{fig:SDcompile}).

\begin{figure}[t]
\centering
\subfigure{
\includegraphics[width=0.46\textwidth]{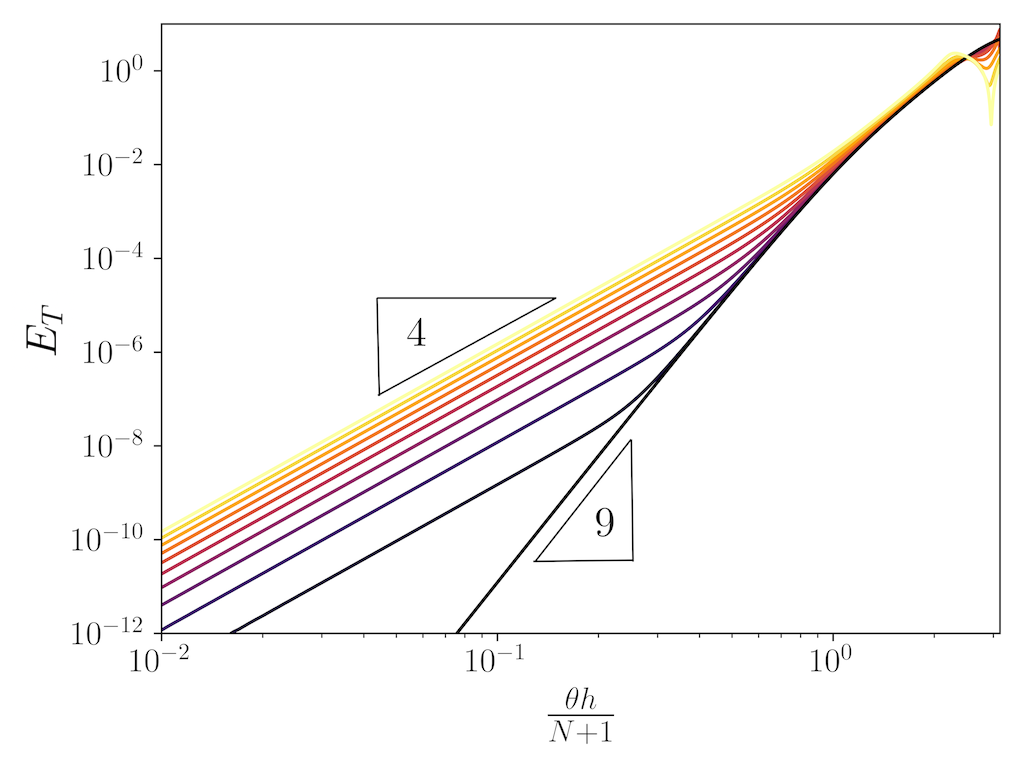}}
\subfigure{
\includegraphics[width=0.46\textwidth]{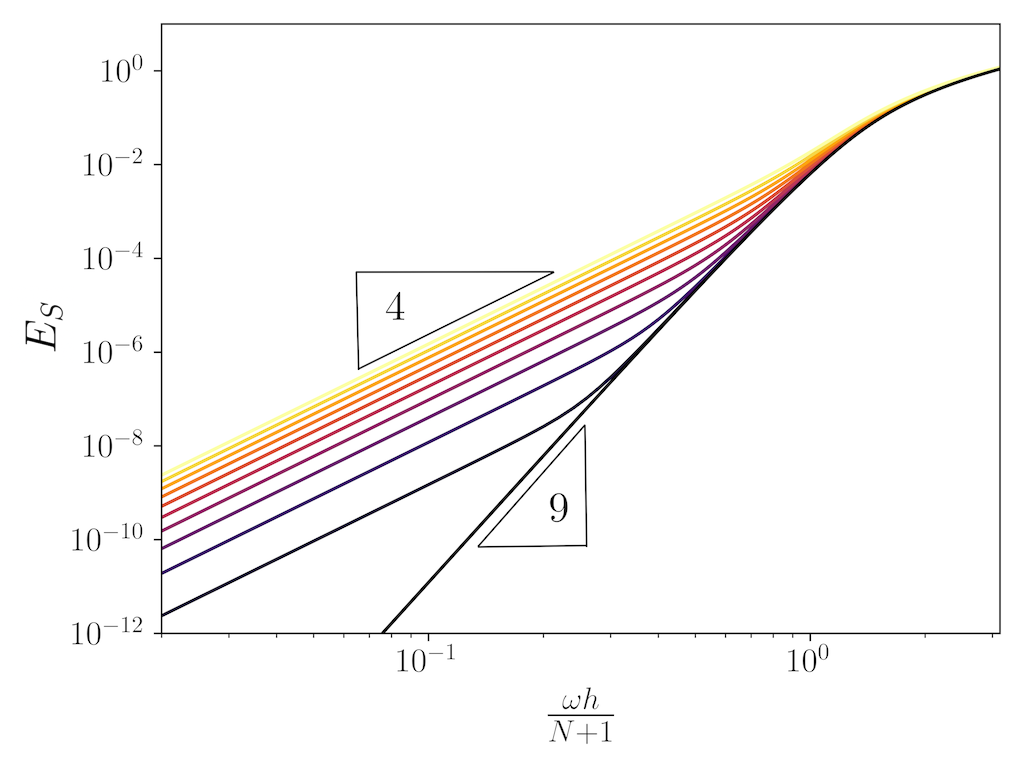}}
\caption{Approximation errors $E_T$ and $E_S$ obtained, respectively, from the temporal (left) and spatial (right) frameworks for a $5^\mathrm{th}$-order FR-SD method coupled with the RK33 scheme.}
\label{SD_k4_RK33_conv}
\end{figure}

To quantify the mutual influence between time and spatial discretisations, the approximation error can be evaluated separately for both temporal and spatial approaches, since the exact solution is known. In particular, for the temporal approach, the error $E_T$ will read
\begin{equation}
E_{T} = | \tilde{\omega}(\theta) - \omega(\theta)|, 
\end{equation}
whereas, in the spatial analysis the error $E_S$ can be evaluated as
\begin{equation}
E_{S} = | \tilde{\theta}(\omega) - \theta(\omega)|.
\end{equation}
Notice, of course, that the estimates above take into account both dispersive and diffusive errors. Both errors have been evaluated for the $5^\mathrm{th}$-order standard upwind FR-SD method coupled with the RK33 scheme. Results are shown in fig.~\ref{SD_k4_RK33_conv}. Similar to what has been found in \cite{vermeire2017behaviour}, for small values of $\theta$, the accuracy of the scheme is reduced by the time integration method. In particular, the expected super-convergence of the FR-SD scheme (of order $2N+1=9$) reduces to a $4^\mathrm{th}$-order convergence. For a certain intermediate range of wavenumbers, instead, the theoretical accuracy of the scheme is preserved. The smaller the prescribed time step, the longer such range will penetrate into the small wavenumbers region. The same applies in the spatial analysis: for small frequencies, the scheme converges with order $4$ and for intermediate frequencies with order $9$. The differences between the two plots are restricted to the high frequencies/wavenumbers region. In fact, since the imaginary parts of both $\tilde{\omega}$ and $\tilde{\theta}$ are negligible for small frequencies/wavenumbers, the temporal and spatial approaches are equivalent, whereby $\tilde{\omega}$ and $\tilde{\theta}$ simply switch axes.

\begin{figure}[t]
\centering
\subfigure{
\includegraphics[width=0.46\textwidth]{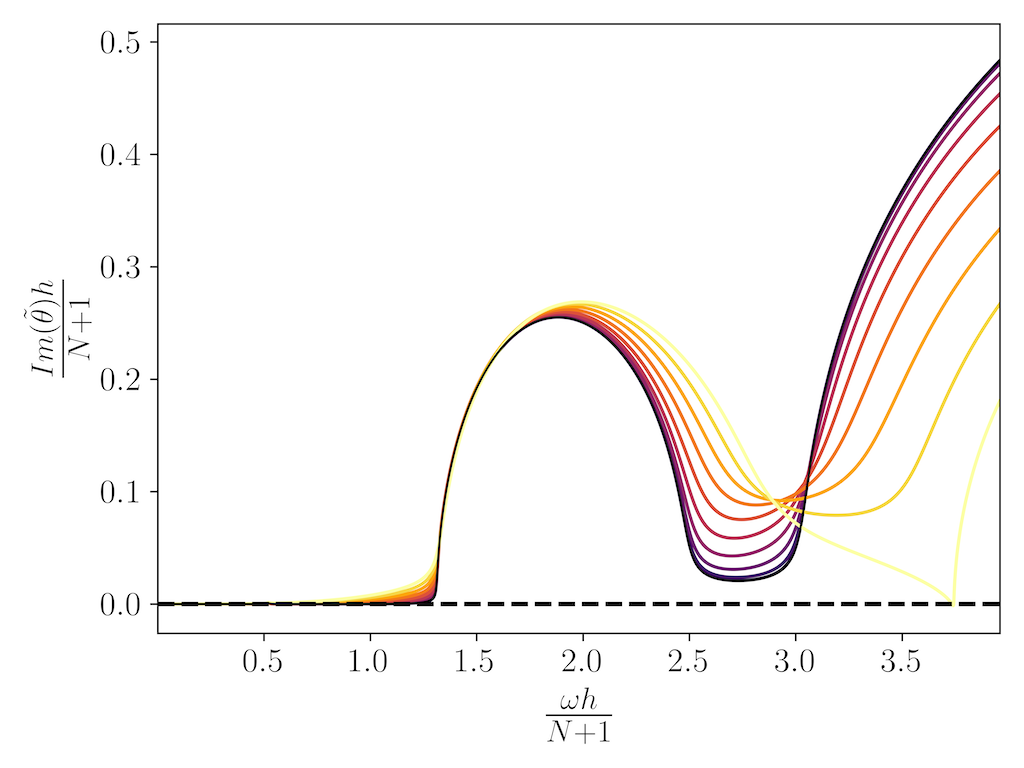}}
\subfigure{
\includegraphics[width=0.46\textwidth]{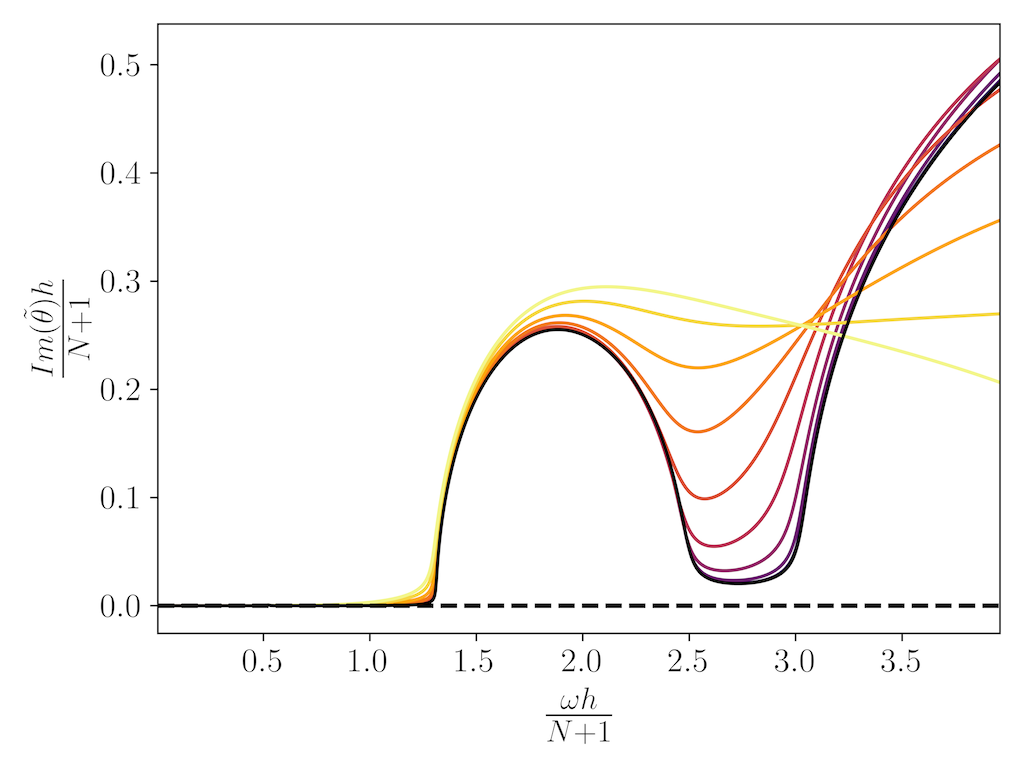}}
\caption{Dissipation curves of the physical transmitted mode at increasing values of $\tau$ (from $\tau = 0$ in black until $\tau = 1$ in yellow). Results obtained for the $5^\mathrm{th}$-order FR-SD with nearly centered flux ($\alpha=0.49$) coupled with either RK33 (left) or RK45 (right).}
\label{SD_k4_RK_cent}
\end{figure}

As already discussed in previous works on spatial eigenanalysis \cite{hu2002eigensolution,mengaldo2018spatial}, whenever the numerical interface flux departs from the standard upwind condition, twice as many roots will exist in the relevant characteristic polynomial. The new roots are typically found to represent wave-like modes that propagate in the opposite direction (reflected modes) and are consequently considered spurious. As already shown for upwind fluxes, for any given value of $\omega$, the number of roots depends on the number of stages of the time integration scheme. In the same way, for non-upwind schemes, the total number of roots is expected to be twice the number of stages. In the Appendix (cf.\ figs.~\ref{fig:SDk2cent} and \ref{fig:SDk4cent}), plots showing all such eigencurves are given. Here, to avoid unnecessary confusion, only the physical modes are shown. The case considered is that of nearly centered flux ($\alpha=0.49$), whose dissipation curves are shown in fig.~\ref{SD_k4_RK_cent}, again for the $5^\mathrm{th}$-order FR-SD method coupled with either RK33 or RK45, as previously shown for the standard upwind case in fig.~\ref{SD_k4_RK}.

Similarly to the standard upwind case, for increasing values of $\tau$, numerical dissipation decreases in the high frequency region. For sufficiently high values of $\tau$, the curve for $\mathrm{Im}(\tilde{\theta})$ is expected to \emph{touch} the horizontal axis with the potential consequent onset of anti-dissipative behaviours, leading to convective instability. This is clearly visible in fig.~\ref{SD_k4_RK_cent} for the RK33 scheme (left). Less so for the RK45 scheme, for which a similar behaviour is expected for high frequencies which are well beyond the plotted range. 
For almost centered fluxes, in the spatial analysis framework, the presence of \emph{dissipative bubbles} has been already observed in previous works \cite{mengaldo2018spatial,tonicello2021comparative}. In other words, significant dissipation appears concentrated in certain frequency ranges, whereas other ranges can be almost completely unaffected. In fig.~\ref{SD_k4_RK_cent}, it can be noticed that, for increasing values of $\tau$, the separation of scales is less strong and the dissipative curves are characterised by a more monotonic behaviour. Non-monotonic dissipation profiles in frequency space might potentially lead to  undesirable non-smooth features in simulations with broadband energy content. It is expected that having neighbouring frequencies so differently affected can deteriorate the distribution of dissipation among scales and impact accuracy and stability of the numerical simulation.

Consequently, only well-resolved wavenumbers/frequencies behave quite similarly in temporal/spatial approaches, whereas large differences are observed in the high wavenumbers/frequencies region. For practical turbulent simulations using high-order discontinuous spectral element methods, depending on the local level of resolution, different ranges of scales and frequencies are of particular interest. For highly resolved flows, such as in DNS, the correlation with spectral analysis will be mainly restricted to the well-resolved regions (i.e.\ small wavenumbers/frequencies). Instead, reducing the resolution of the scheme and passing from DNS to LES of turbulent flows, higher wavenumber and frequency regions in the plots will provide more useful information.

\section{Numerical experiments} \label{S:5}
In order to validate the theoretical framework introduced in previous sections, a series of numerical experiments is presented to quantify the interplay between temporal and spatial errors for practical flows. First, we consider the discretisation of the one-dimensional linear advection using the SD scheme. We then take into account more complex numerical simulations of the Euler equations in order to evaluate the influence of the time step on both well-resolved and under-resolved flows. In all the Euler computations presented herein, if not stated differently, the original Roe flux has been employed as a representative of the standard upwind flux.
All the simulations presented in the following subsections have been performed using the improved SD method by Liang and Jameson~\cite{liang2009high,liang2009spectral,jameson:10}.
\subsection{One-dimensional linear advection equation}
\begin{figure}[t]
\centering
\subfigure[RK33]{
\includegraphics[width=0.46\textwidth]{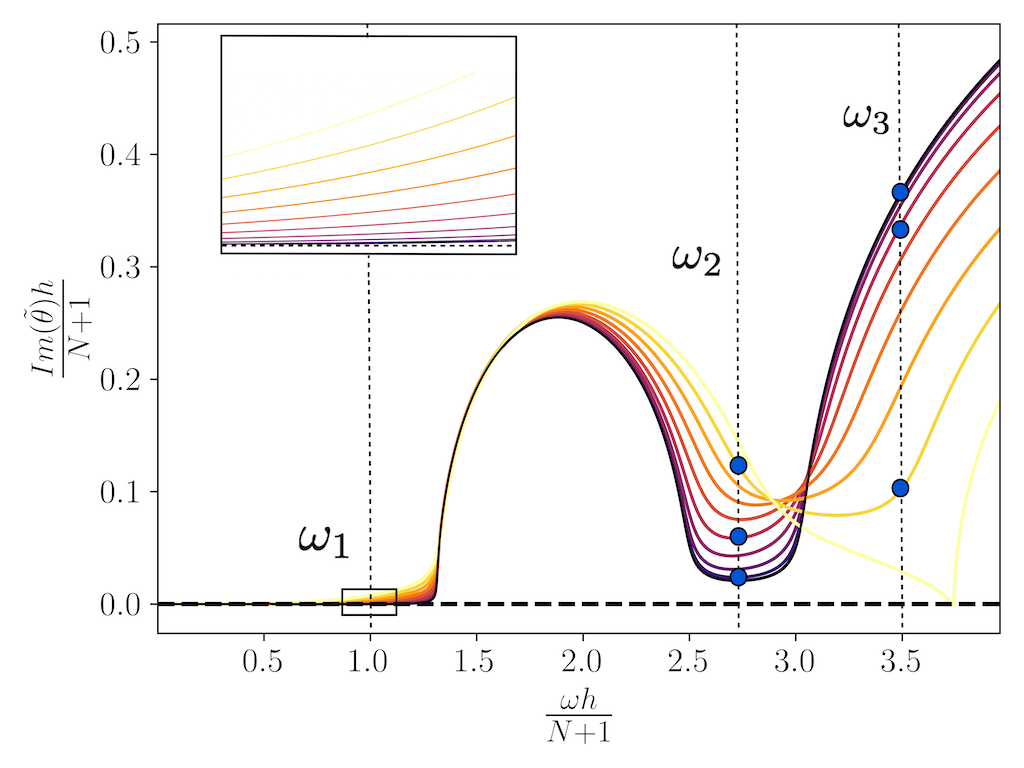}}
\subfigure[RK45]{
\includegraphics[width=0.46\textwidth]{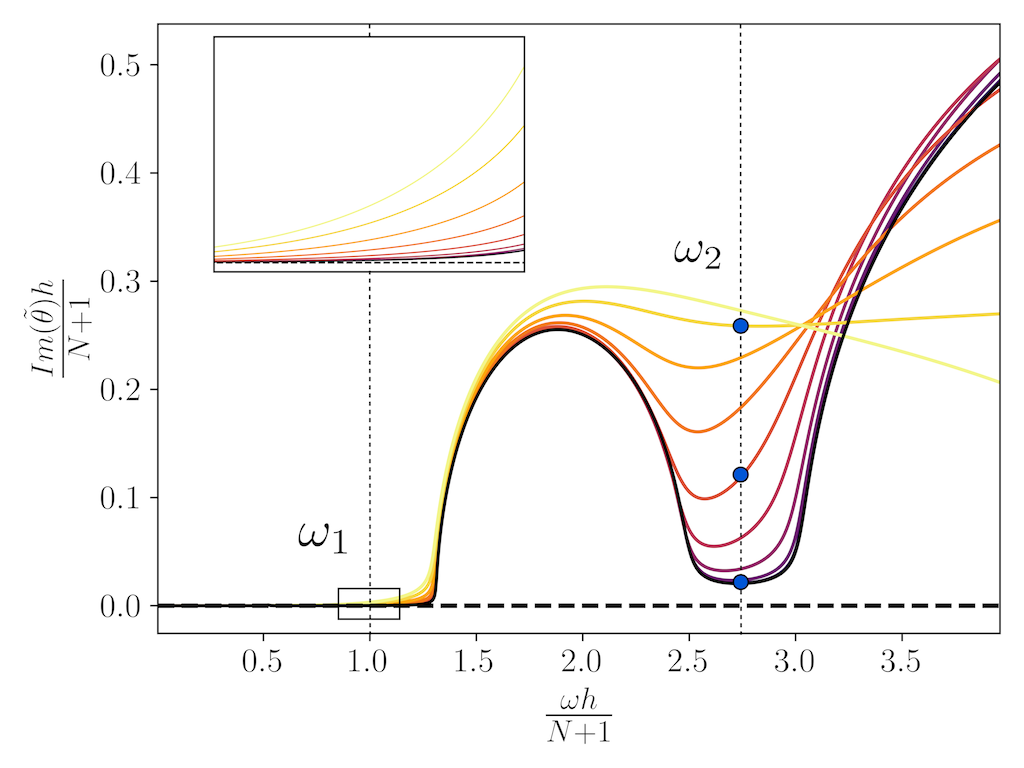}}
\caption{Physical dissipative curves varying $\tau$ (FR-SD) using almost centered numerical fluxes. Dashed, vertical lines denote the frequencies $\omega_{1}$, $\omega_{2}$ and $\omega_{3}$ (only for RK33). For $\omega_{2}$ and $\omega_{3}$ the dots indicate the levels of dissipation expected for the specific values of $\tau$ used in the simulations.}
\label{SD_k4_RK_cent2}
\end{figure}

The first numerical test consists in the discretisation of the one-dimensional linear advection equation with an oscillating inlet boundary condition. 
This simulation allows to monitor the general behaviour of dissipative bubbles when the prescribed time step approaches the CFL limit.
Tests are conducted using almost centered numerical fluxes and RK33 or RK45 time integration schemes for different injected frequencies.
Accordingly, fig.~\ref{SD_k4_RK_cent2} shows the same curves presented in fig.~\ref{SD_k4_RK_cent} in which the  tested frequencies are indicated in order to appreciate the expected level of dissipation of each test.
The relevant  equivalent frequencies are $\omega_{1} = 1$, $\omega_{2} = 2.75$ and $\omega_{3} = 3.5$ (only for the RK33 test), which are highlighted in fig.~\ref{SD_k4_RK_cent2} using dashed vertical lines. Close to $\omega_{2}$ and $\omega_{3}$, the expected levels of numerical dissipation are highlighted using blue dots.
%

\begin{figure}[t]
\centering
\subfigure[$\tau=0.1$]{
\includegraphics[width=0.31\textwidth]{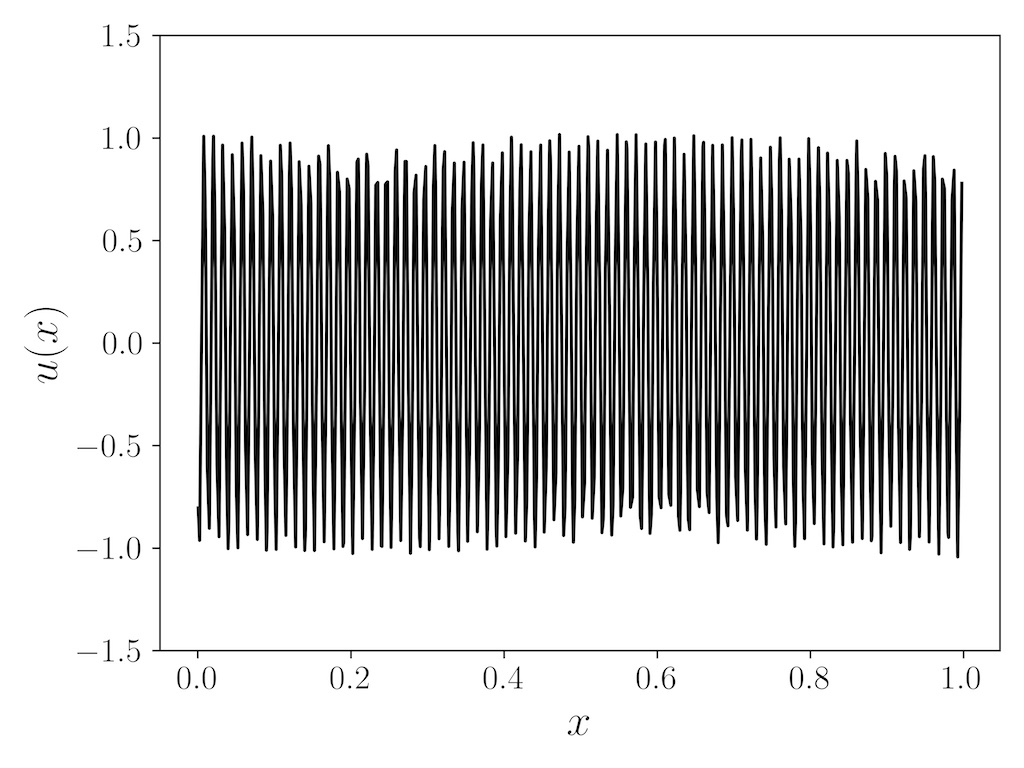}}
\subfigure[$\tau =0.5$]{
\includegraphics[width=0.31\textwidth]{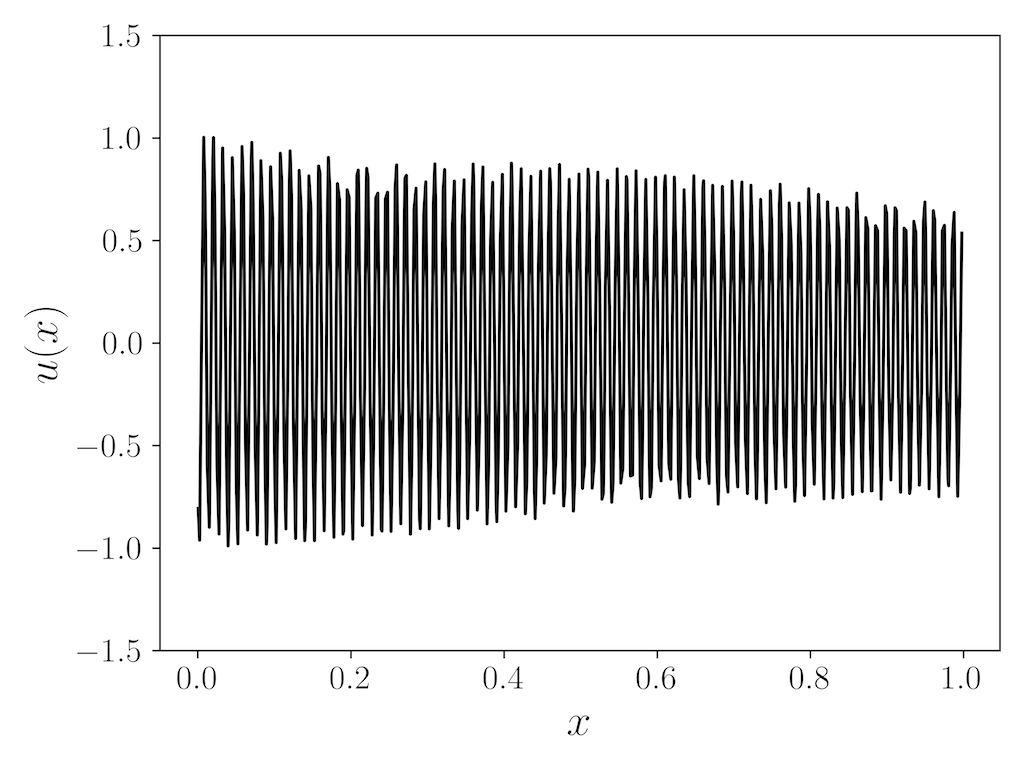}}
\subfigure[$\tau =0.9$]{
\includegraphics[width=0.31\textwidth]{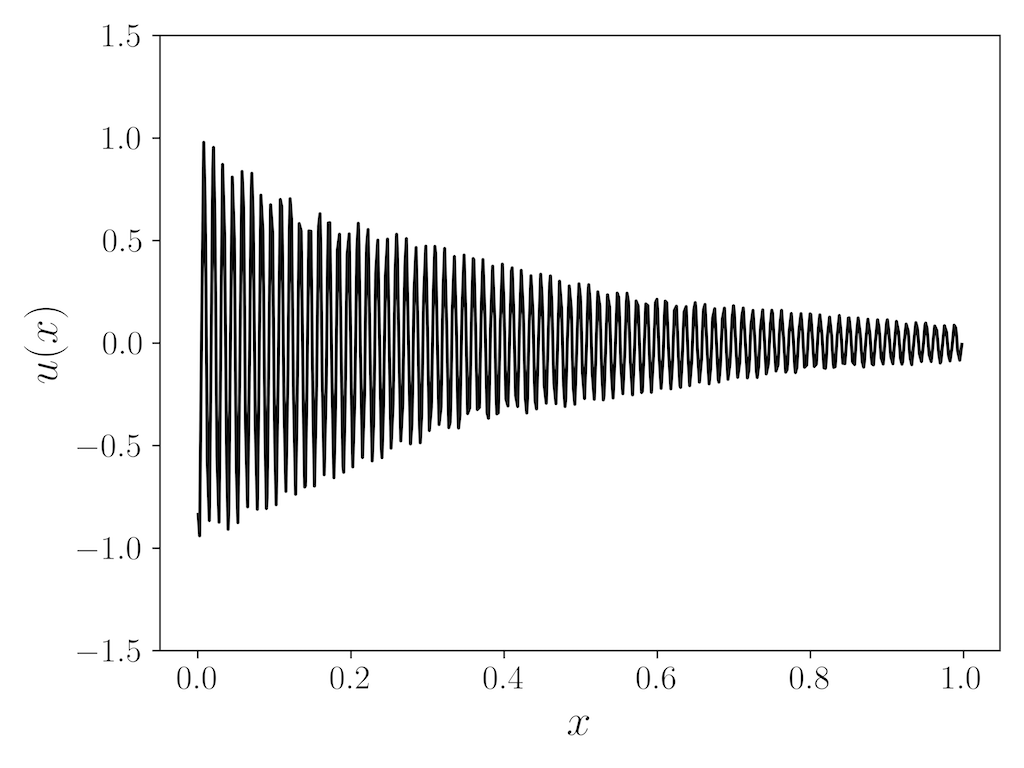}}
\caption{Solution signal $u(x)$ at prescribed frequency $\omega_{1}$ for different values of $\tau$ (RK33).}
\label{RK33_w500}
\end{figure}

\begin{figure}[t]
\centering
\subfigure[$\tau=0.1$]{
\includegraphics[width=0.31\textwidth]{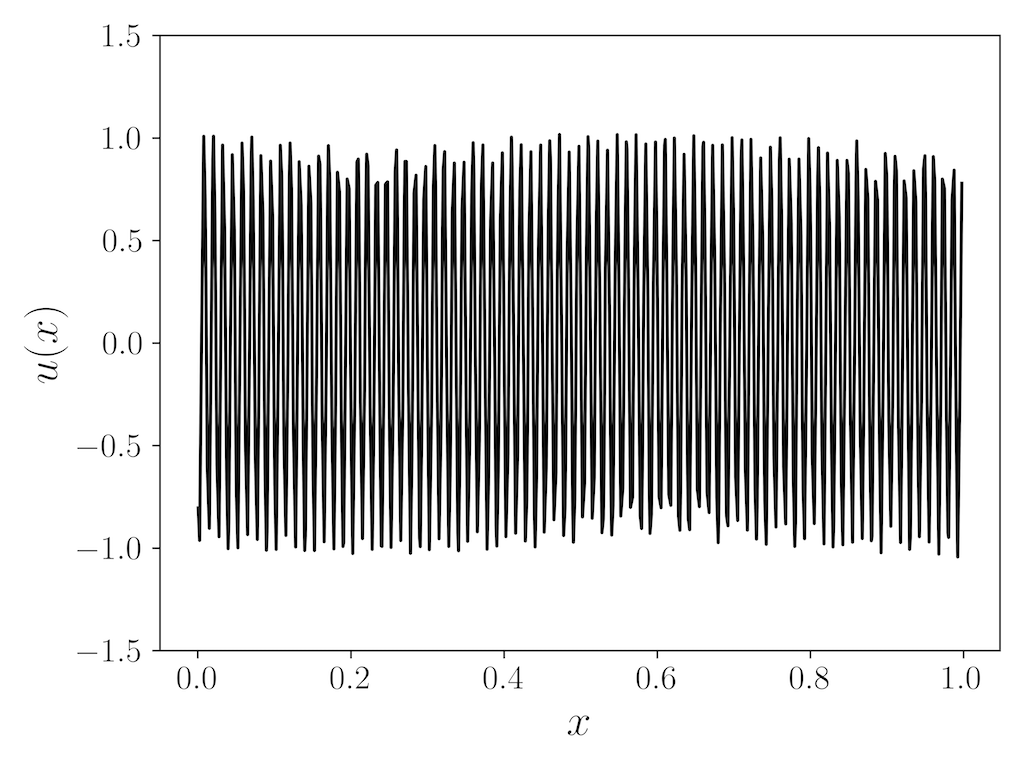}}
\subfigure[$\tau =0.5$]{
\includegraphics[width=0.31\textwidth]{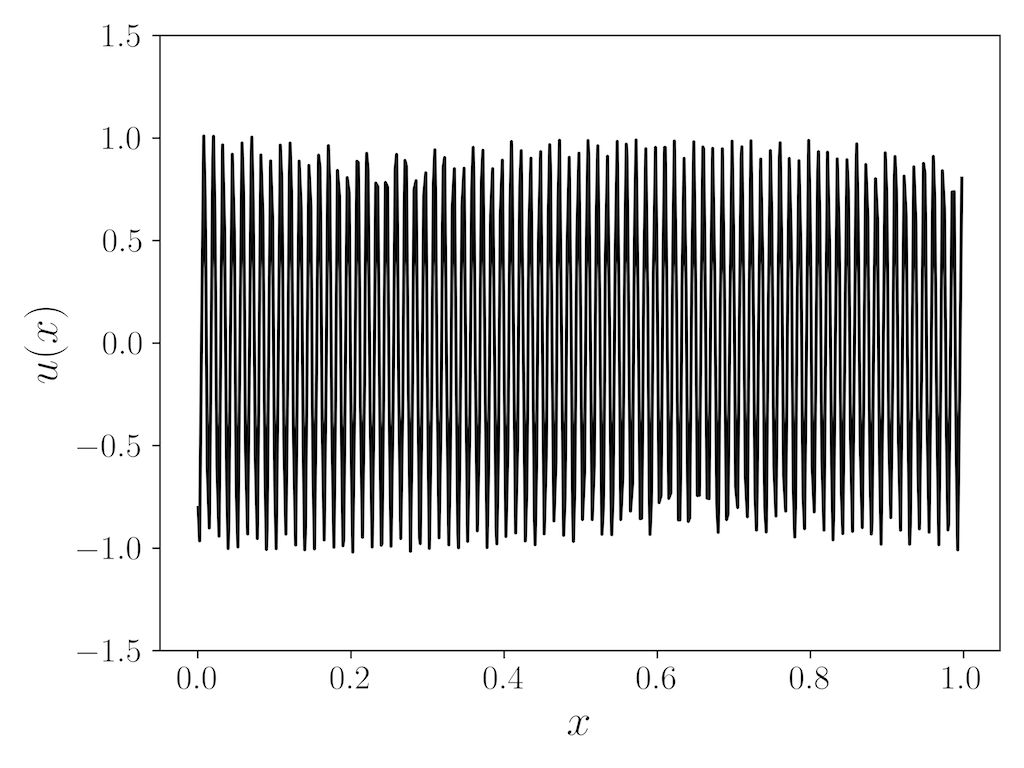}}
\subfigure[$\tau =0.9$]{
\includegraphics[width=0.31\textwidth]{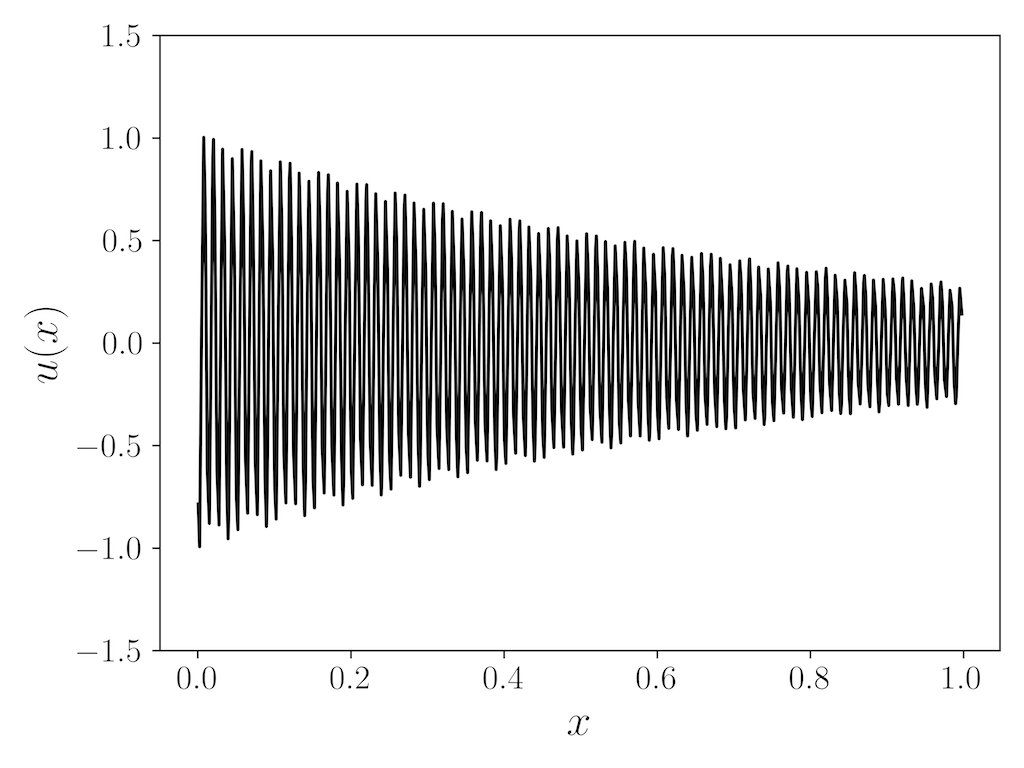}}
\caption{Solution signal $u(x)$ at prescribed frequency $\omega_{1}$ for different values of $\tau$ (RK45).}
\label{RK45_w500}
\end{figure}

\begin{figure}[t]
\centering
\subfigure[$\tau=0.1$]{
\includegraphics[width=0.31\textwidth]{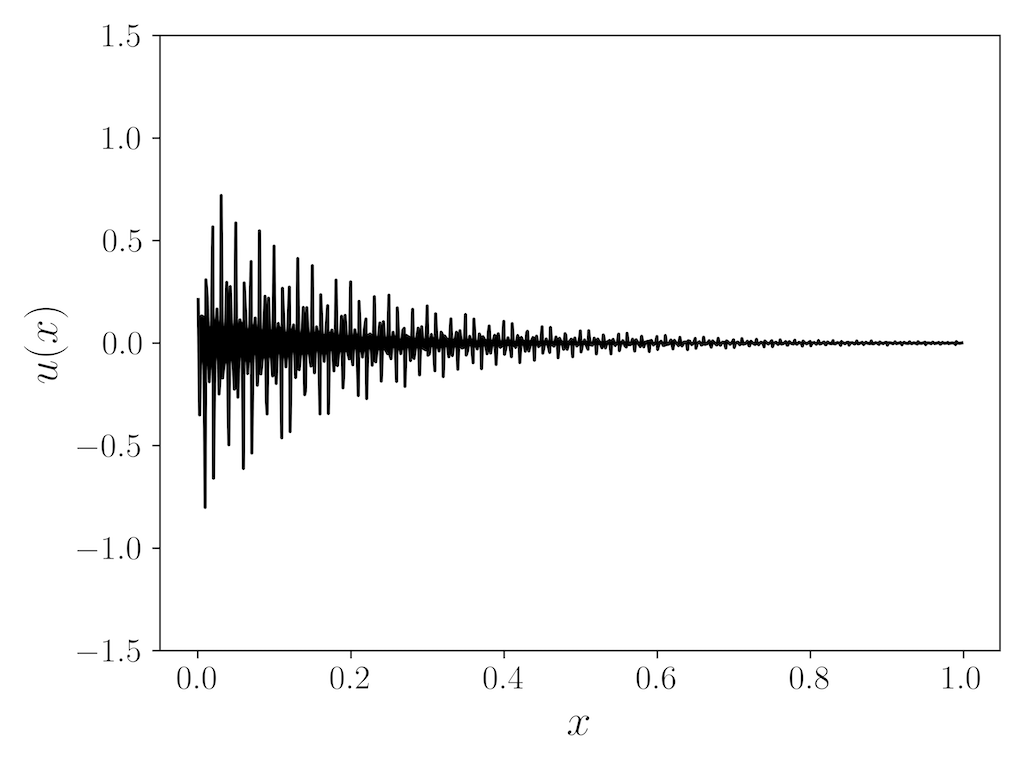}}
\subfigure[$\tau =0.5$]{
\includegraphics[width=0.31\textwidth]{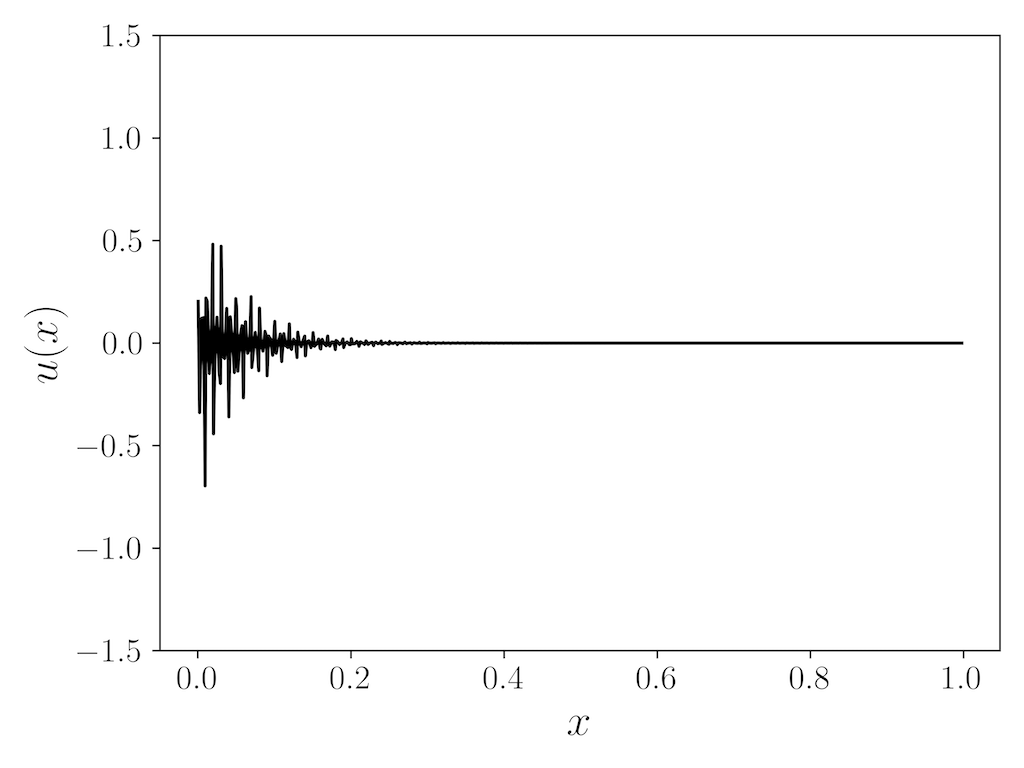}}
\subfigure[$\tau =0.9$]{
\includegraphics[width=0.31\textwidth]{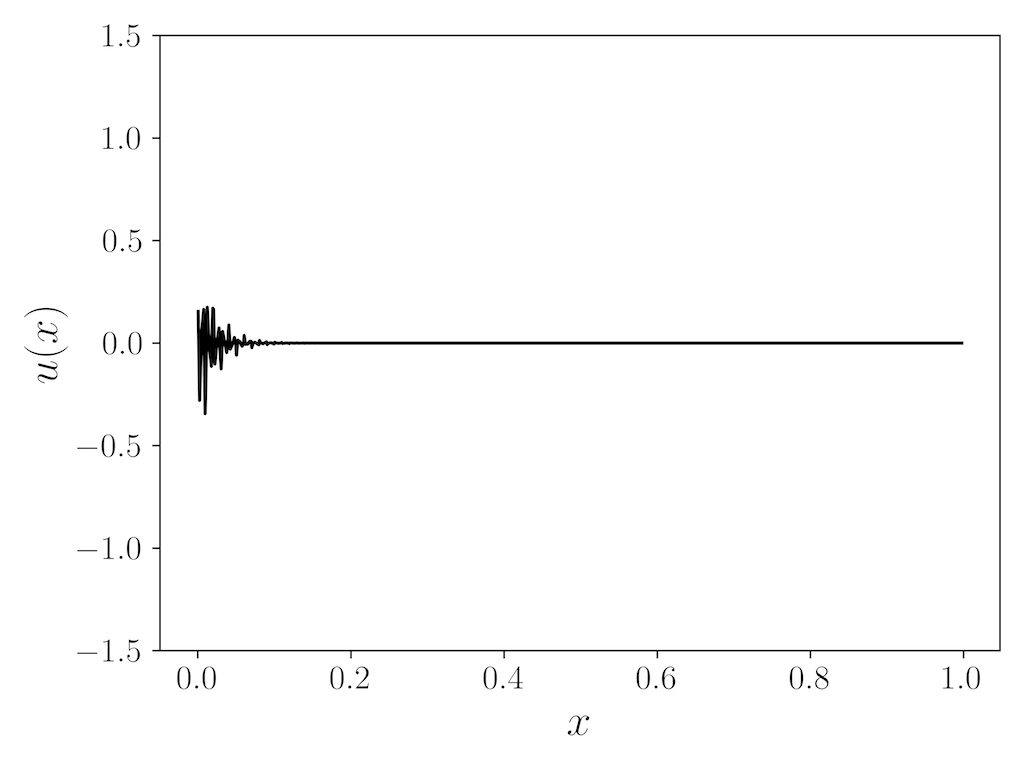}}
\caption{Solution signal $u(x)$  at prescribed frequency $\omega_{2}$ for different values of $\tau$ (RK33).}
\label{RK33_w1375}
\end{figure}

\begin{figure}[t]
\centering
\subfigure[$\tau=0.1$]{
\includegraphics[width=0.31\textwidth]{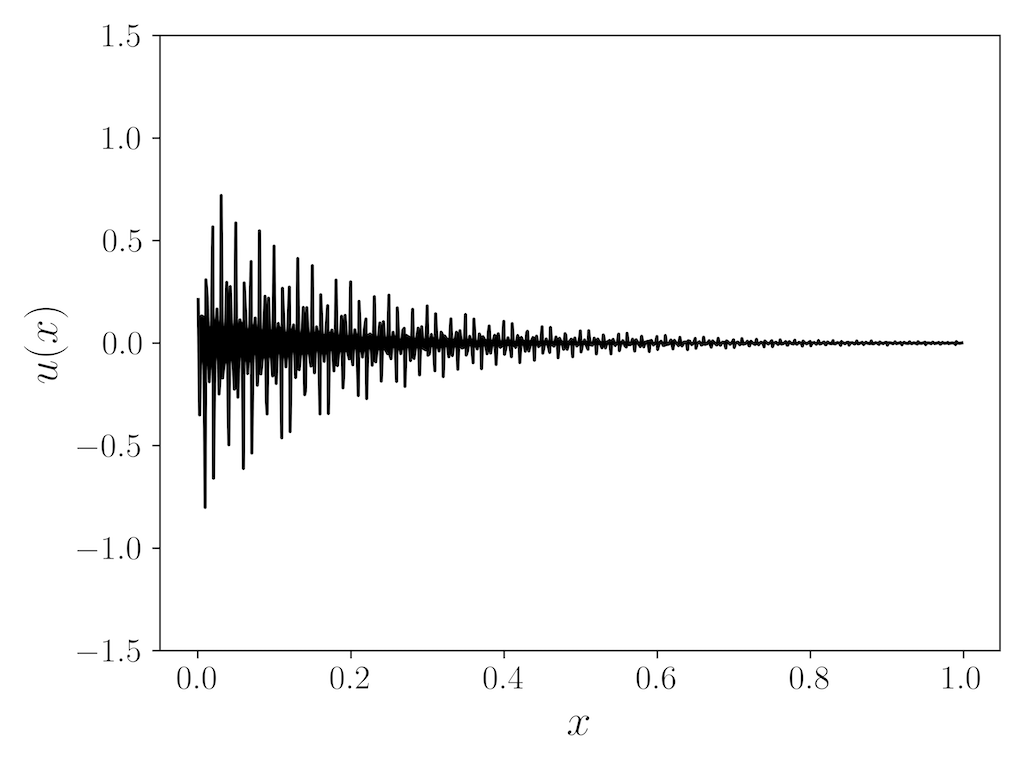}}
\subfigure[$\tau =0.5$]{
\includegraphics[width=0.31\textwidth]{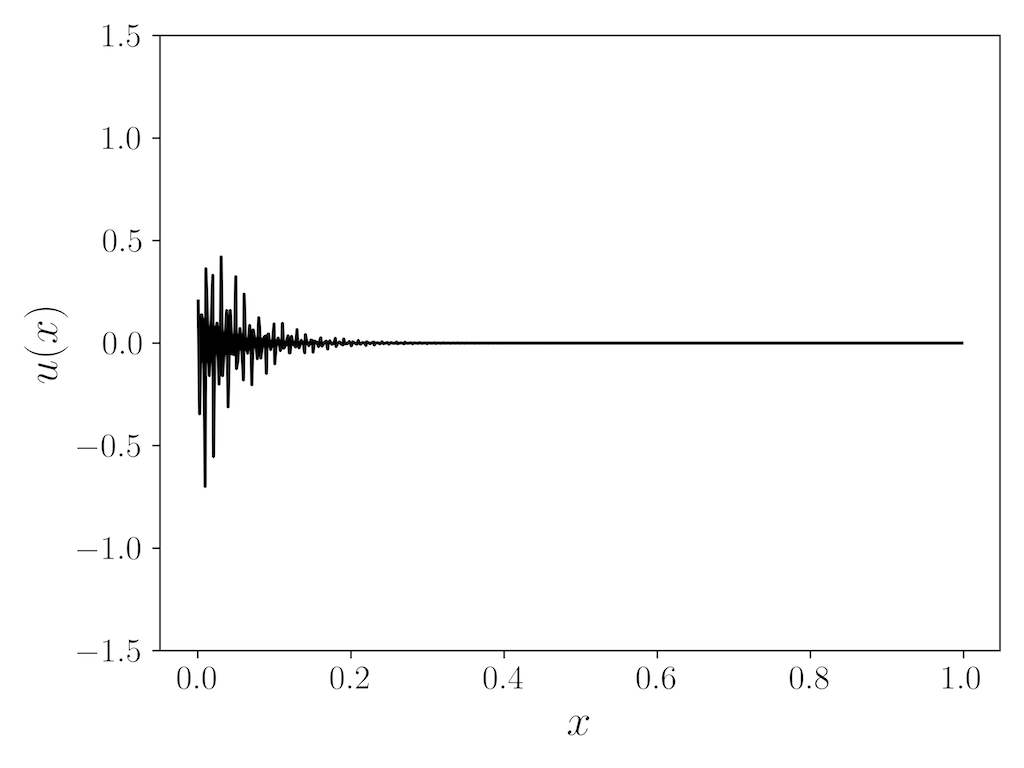}}
\subfigure[$\tau =0.9$]{
\includegraphics[width=0.31\textwidth]{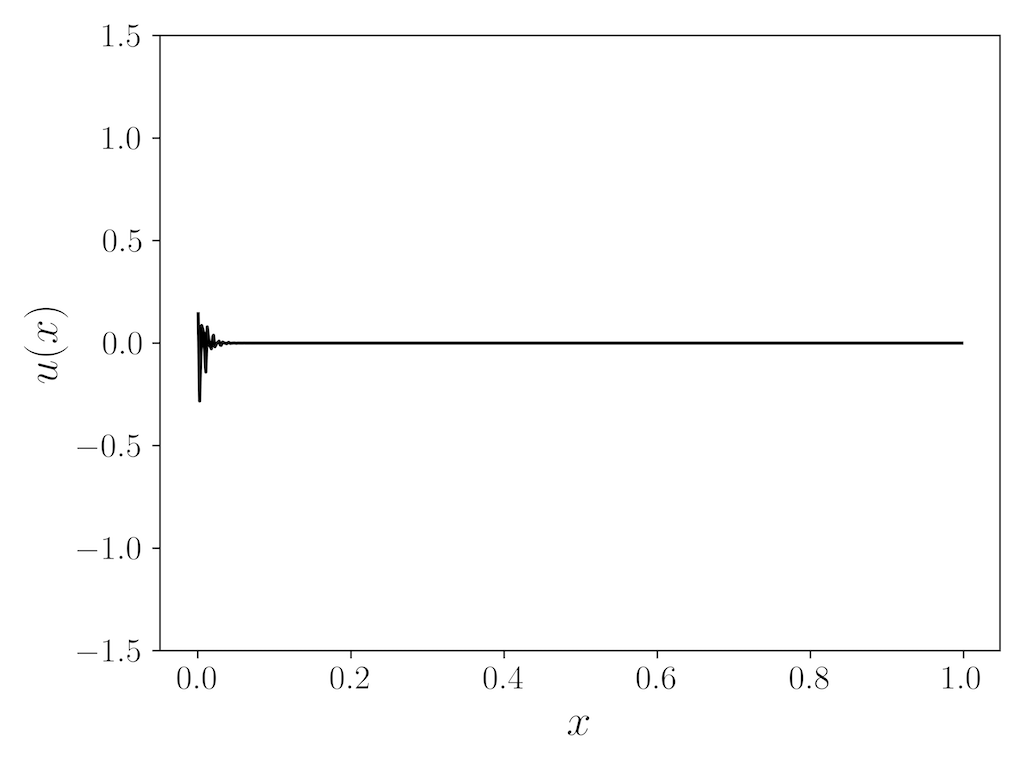}}
\caption{Solution signal $u(x)$ at prescribed frequency $\omega_{2}$ for different values of $\tau$ (RK45).}
\label{RK45_w1375}
\end{figure}

The domain $\Omega = [0,1]$ is split into $100$ equally spaced elements ($h=0.01$). 
The frequency of the inlet signal can then be directly linked to the equivalent frequency $\omega h/(N+1)$. The simulation matches the theoretical analysis: a $5^{\mathrm{th}}$ order SD discretisation, coupled with a RK33/RK45 time integration schemes has been employed. To see the influence of the temporal errors, increasing values of $\tau$ have been considered. 
First, in fig.~\ref{RK33_w500}, the velocity signal $u(x)$ is plotted along the $x$ direction for $\omega=\omega_{1}$ and for different values of $\tau$ using a RK33 time marching scheme. According to the theoretical analysis presented in the first part of the paper, for increasing values of $\tau$, larger values of dissipation are expected (see fig.~\ref{SD_k4_RK_cent2}). Such conclusion is confirmed by the numerical simulations shown in fig.~\ref{RK33_w500}. The amplitude of the signal $u(x)$ tends, in fact, to decrease for larger time steps. 
The same experiment has been repeated using a RK45 scheme and the results are shown in fig.~\ref{RK45_w500}. The same conclusions apply: increasing the time step magnitude leads to an increase of numerical dissipation. Notice, furthermore, that for $\tau=0.9$ the signal is more dissipated using the RK33 scheme rather than the RK45 one. In fact, from fig.~\ref{SD_k4_RK_cent2}, it can be noticed that the dissipative curves, in proximity of $\omega_{1}$, tend to increase a bit more using the RK33 scheme rather than the RK45 one. 
The same comparison has been performed for the frequency $\omega_{2}$, which is located in between dissipative bubbles. More or less the same behaviour is expected, where larger values of $\tau$ lead to stronger numerical dissipation. The solution $u(x)$ is shown for the RK33 scheme in fig.~\ref{RK33_w1375} and for the RK45 scheme in fig.~\ref{RK45_w1375}. The only interesting difference is that, in agreement with the theoretical analysis, for $\tau =0.9$ the RK33 scheme is less dissipative than the RK45. For $\tau \approx 0$ and $\tau=0.5$, instead, the levels of numerical dissipation are quite similar. 

\begin{figure}[t]
\centering
\subfigure[$\tau=0.1$]{
\includegraphics[width=0.31\textwidth]{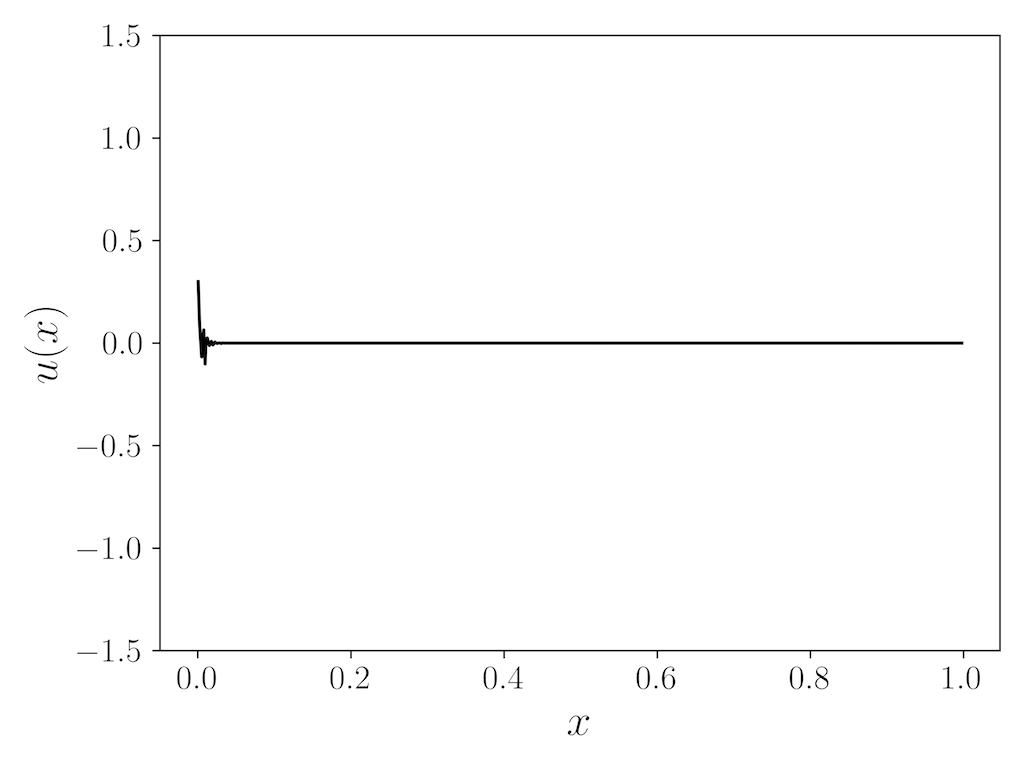}}
\subfigure[$\tau =0.5$]{
\includegraphics[width=0.31\textwidth]{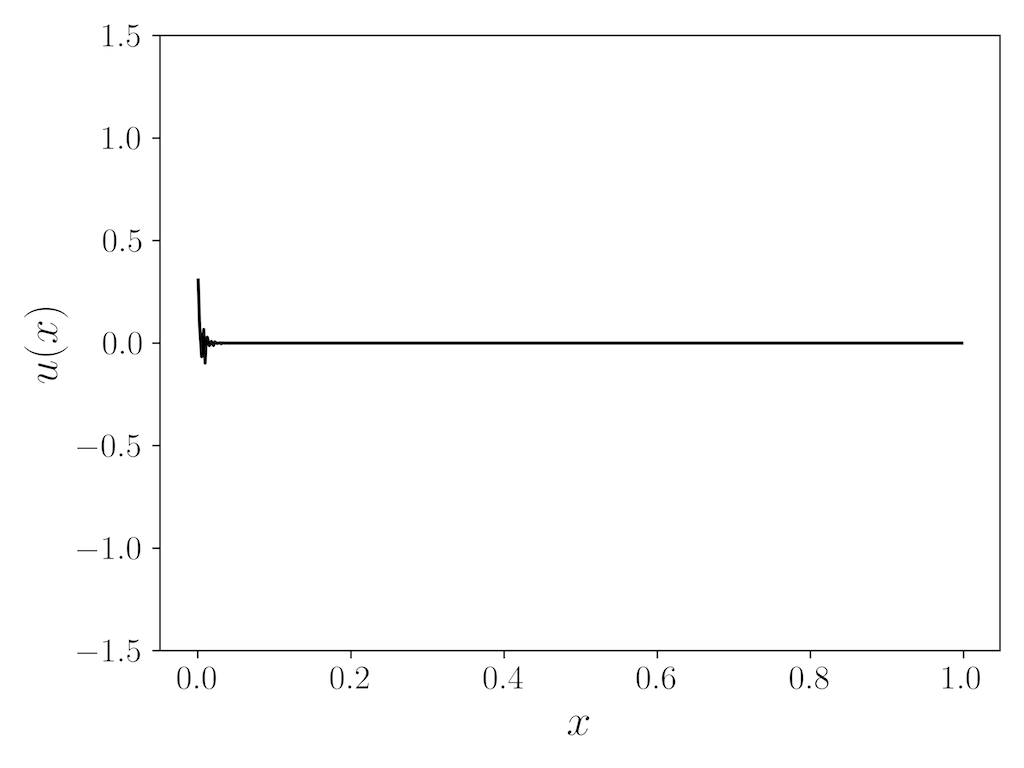}}
\subfigure[$\tau =0.9$]{
\includegraphics[width=0.31\textwidth]{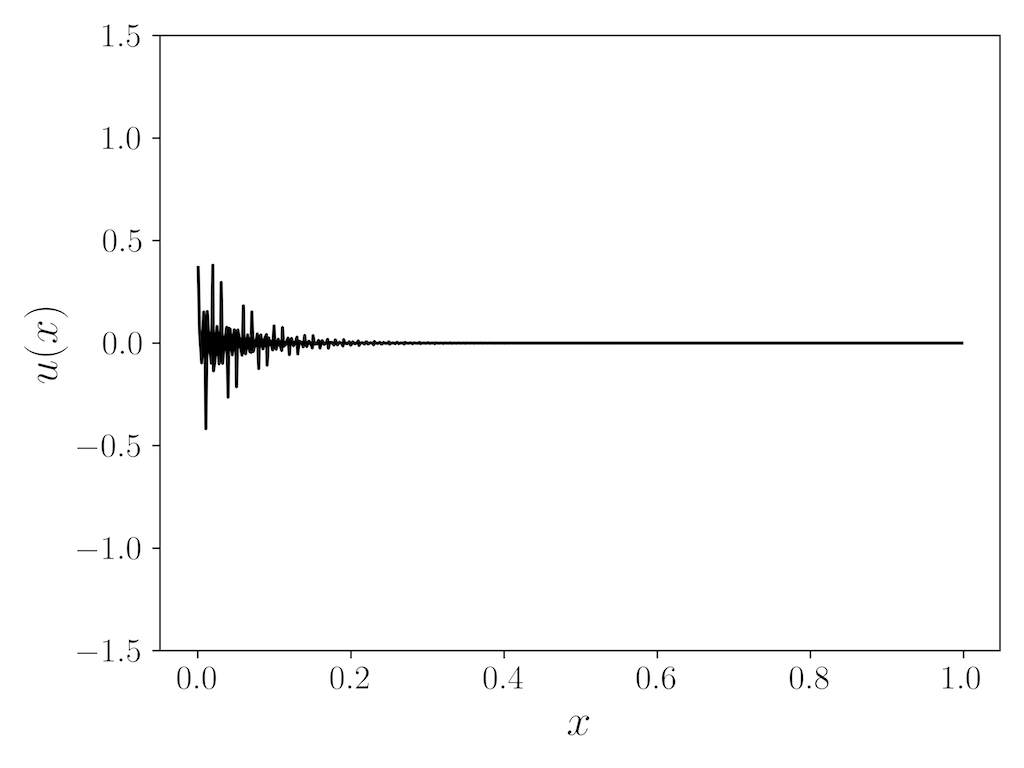}}
\caption{Solution signal $u(x)$ at prescribed frequency $\omega_{3}$ for different values of $\tau$ (RK33).}
\label{RK33_w1750}
\end{figure}

In the RK33 case, the additional frequency $\omega_{3}$ has been considered. At this frequency, in fact, a different behaviour with respect to the low frequencies is expected: for increasing values of $\tau$, numerical dissipation is expected to decrease instead of increase like for all the other frequencies. Such behaviour is confirmed in fig.~\ref{RK33_w1750} where for $\tau$ very close to the CFL limit the numerical dissipation is significantly lower. In similarity with fully-discrete temporal eigenanalysis, the highest frequencies are the first to experience instability. Then, in a general sense, the present test case suggests that for a reasonably large, low frequency region, the influence of the time integration scheme is essentially dissipative and larger time steps cause stronger dissipation. In the high-frequency region, instead, the behaviour is the opposite: in the semi-discrete analysis such frequencies are usually dumped very fast, whereas, approaching the CFL limit, numerical dissipation decrease significantly, until, eventually, it changes sign leading the simulation to be unstable. Consequently, in similarity with respect the fully-discrete temporal eigenanalysis, the largest frequencies are responsible for the CFL instability. 

\begin{figure}[t]
\centering
\subfigure{\includegraphics[width=0.46\textwidth]{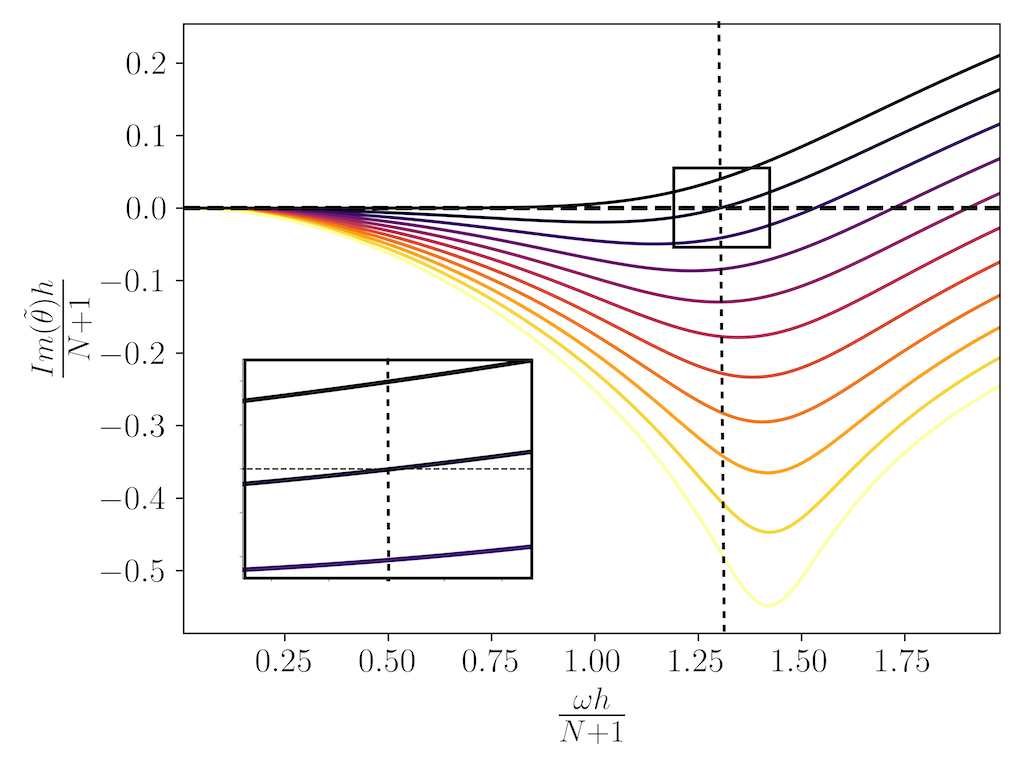}}
\subfigure{\includegraphics[width=0.46\textwidth]{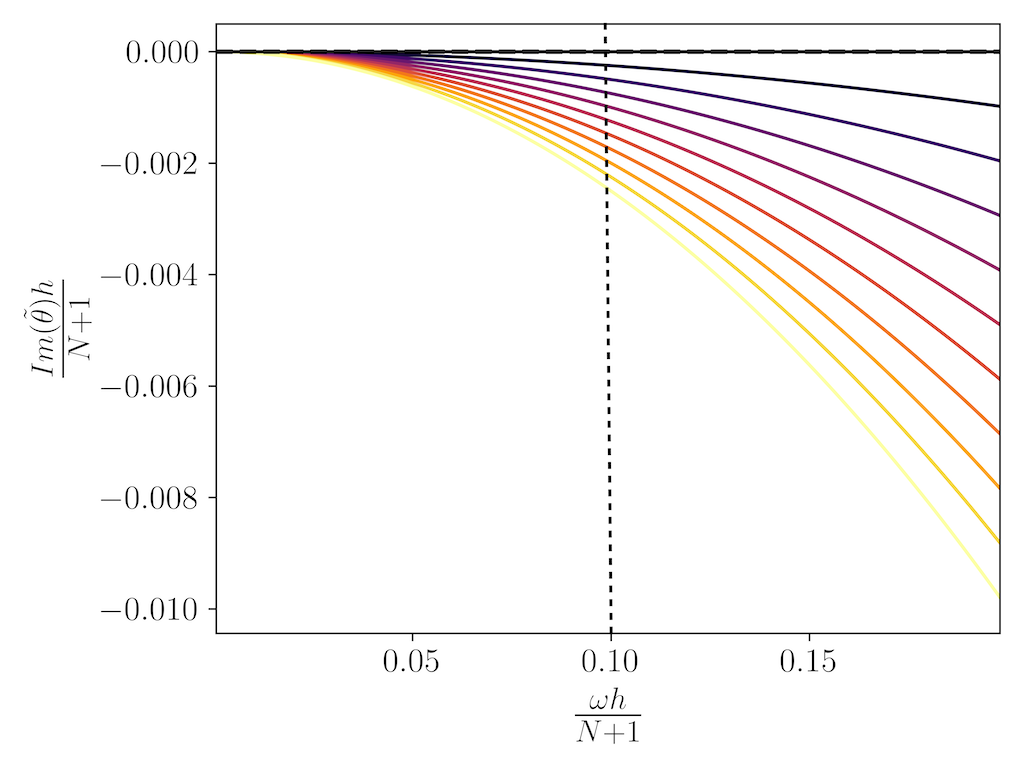}}
\caption{Physical dissipative curves varying $\tau$ (FR-SD) using Euler time-integration. On the left, a closer look near the origin. Vertical lines indicate the frequencies that have been numerically simulated.}
\label{fig:SD_k4_RK11_adv}
\end{figure}

Finally it is interesting to study the behaviour of a scheme that, according to the fully-discrete spatial eigenanalysis herein presented, is expected to be unstable for any choice of time-step. The same computational grid and order of approximation have been chosen. In similarity with previous tests, a series of frequencies have been simulated. In fig.~\ref{fig:SD_k4_RK11_adv} the dissipative curves for a $5^{\mathrm{th}}$ order FR-SD discretisation coupled with an explicit Euler time-integration is shown. In particular, in order to quantify the instability of such space/time discretisation, the following equivalent frequencies have been considered: $\omega_{0}=0.1$ and $\omega_{1}=1.3$. The low frequency test is meant to investigate the instability of the prescribed discretisation for any choice of computational time step. We expect that for any choice of time-step the prescribed sinusoidal signal will grow while propagating in the domain. For very low frequencies the increase will likely be very small, but nonetheless still present.

\begin{figure}[t]
\centering
\includegraphics[width=0.9\textwidth]{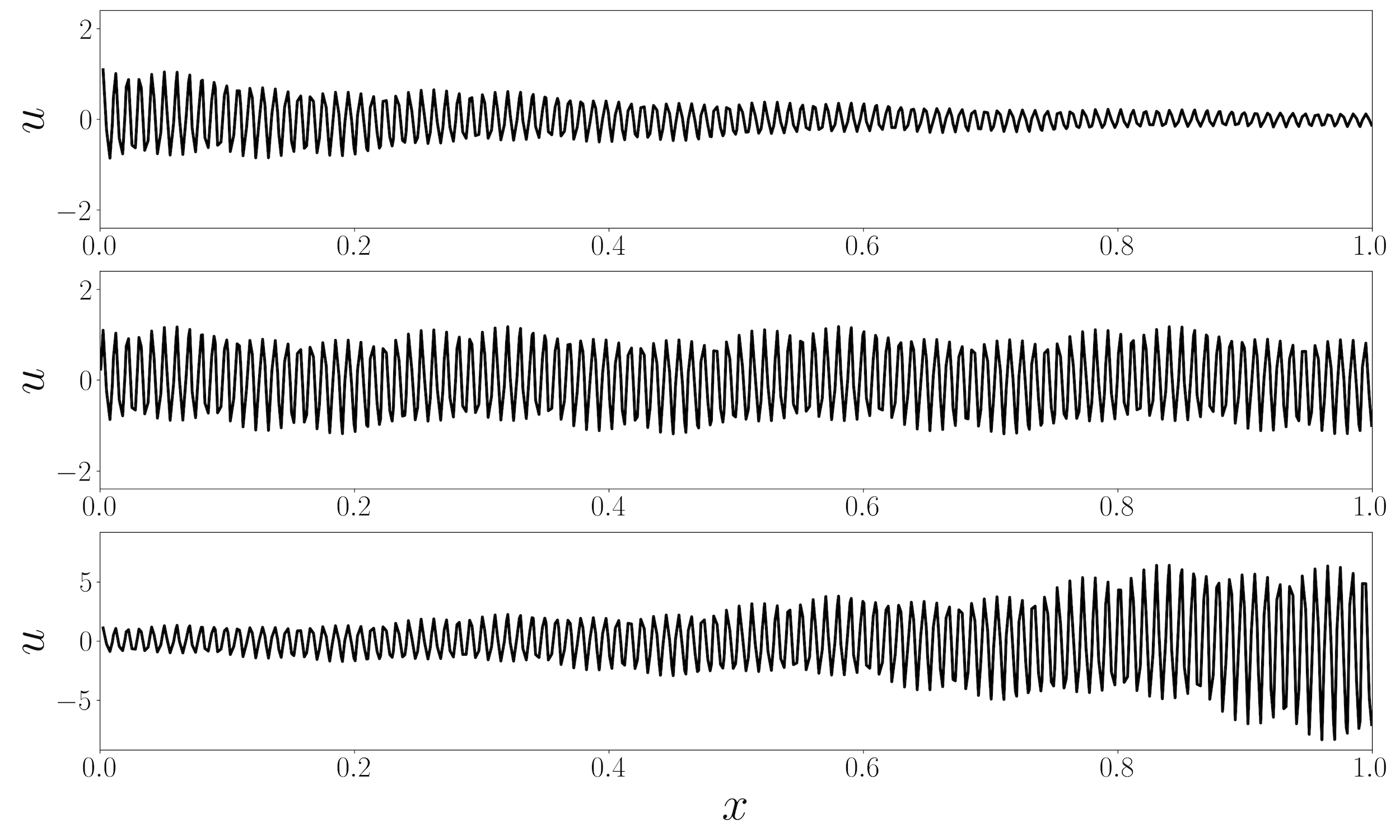}
\caption{Numerical solution of $\omega_{0} = 0.1$ using explicit Euler time integration scheme. Top, $\Delta t = 9.825 \times 10^{-5}$; Center $\Delta t = 9.925 \times 10^{-5}$; Bottom $\Delta t = 1.0925 \times 10^{-4}$.}
\label{fig:zero_diss}
\end{figure}

For the second frequency, instead, a different behaviour is expected to take place: for sufficiently small time steps the numerical scheme should dissipate the inlet oscillations, but, for increasing time steps, numerical dissipation should decrease, until it becomes anti-dissipation, leading the numerical solution to grow while propagating in the domain. The sudden change of behaviour, from dissipative to anti-dissipative is predicted to take place at $\Delta t \approx 0.01$ from the theoretical analysis. The intersection between the dissipative curve for $\Delta t \approx 0.01$ and the $x$-axis is highlighted in fig.~\ref{fig:SD_k4_RK11_adv}. Consequently, after proper rescaling by the grid size of the present simulation ($h=0.01$), we expect to observe such change of trend for time steps around $\Delta t = 1.0 \times 10^{-4}$. Three time-step values are therefore tested near $\Delta t = 1.0 \times 10^{-4}$, as shown in fig.~\ref{fig:zero_diss}.

The numerical solution behaves as expected, where a sudden change from dissipative to anti-dissipative character is observed for increasing time steps. For $\Delta t = 9.925 \times 10^{-5}$, the prescribed signal at the inlet is barely affected by numerical dissipation. For smaller time steps, numerical dissipation increases and the sinusoidal signal is expected to decrease in amplitude as numerically confirmed in the top plot of fig.~\ref{fig:zero_diss}.
Finally, for time steps slightly larger, instead, negative values of the imaginary part of $\tilde{\theta}$ arise and the solution is expected to grow over space, as confirmed in the bottom plot of fig.~\ref{fig:zero_diss}. Numerical results are then in very good agreement with the theoretical analysis herein presented which is able to properly predict the effect of numerical dissipation in a fully-discrete framework where both spatial and temporal errors contribute on the overall accuracy and regularity of the numerical solution.

\begin{figure}[t]
\centering
\includegraphics[width=0.9\textwidth]{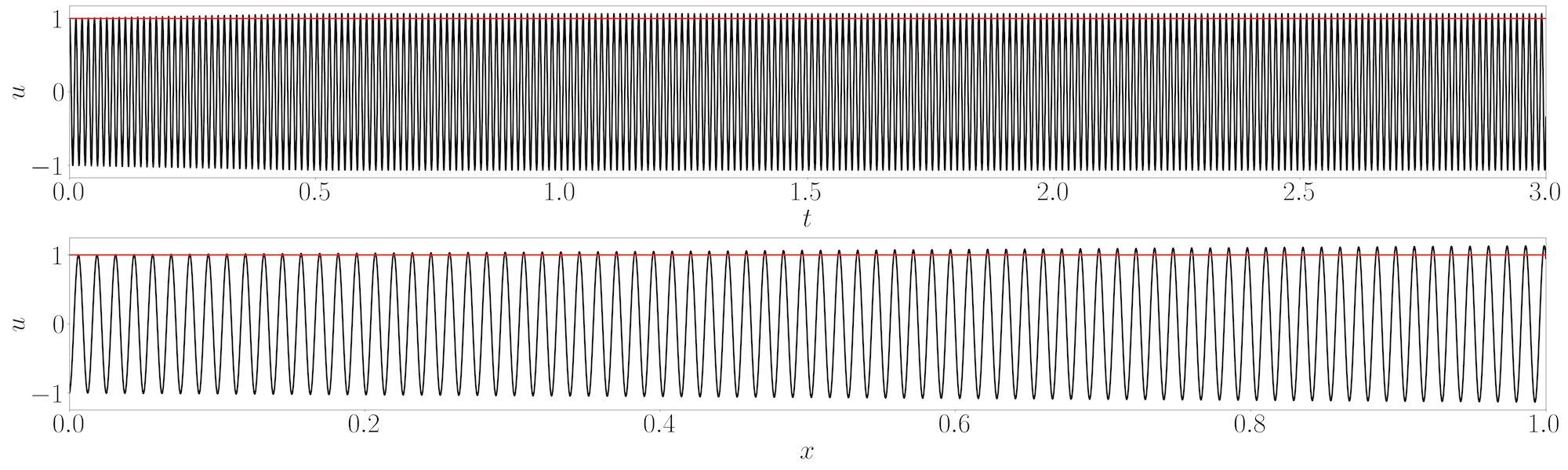}
\caption{Numerical solution of $\omega_{1} = 1.3$ using explicit Euler time integration scheme with $\Delta t = 1.0 \times 10^{-6}$. Top, probed solution at $x=0.5$; Bottom, numerical solution at the end of the simulation. The red horizontal line indicates the iso-line $u=1$.}
\label{fig:w0_small}
\end{figure}

\begin{figure}[t]
\centering
\includegraphics[width=0.9\textwidth]{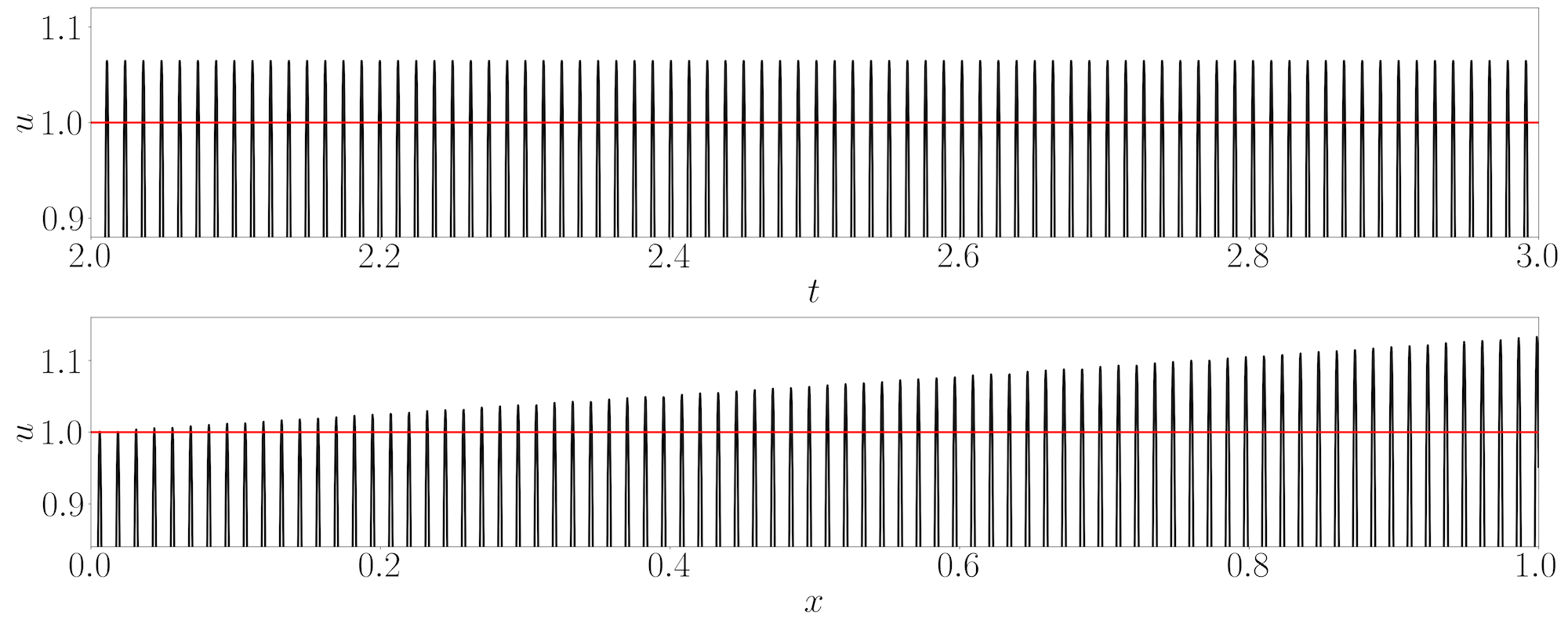}
\caption{Closer look of the solution of fig.~\ref{fig:w0_small} around $u=1$ (red line).}
\label{fig:w0_small_detail}
\end{figure}

Regarding the simulation of the lowest frequency, in order to allow a sufficient large number of periods within the unitary domain, the number of elements has been increased to $1000$. In fig.~\ref{fig:w0_small} the results of the simple advection of the sinusoidal signal with lowest frequency is shown for the explicit Euler scheme with a considerably smaller time step, namely $\Delta t = 1.0 \times 10^{-6} = 5 \times 10^{-3} h/(N+1)$. On the top, the solution is probed at $x=0.5$, on the bottom, instead, the numerical solution at the end of the simulation is shown. As predicted by the theoretical analysis, the solution tends to grow over space. Such behaviour is even more evident in fig.~\ref{fig:w0_small_detail}, where the solution is zoomed close to the line $u=1$.


\begin{figure}[t]
\centering
\includegraphics[width=0.9\textwidth]{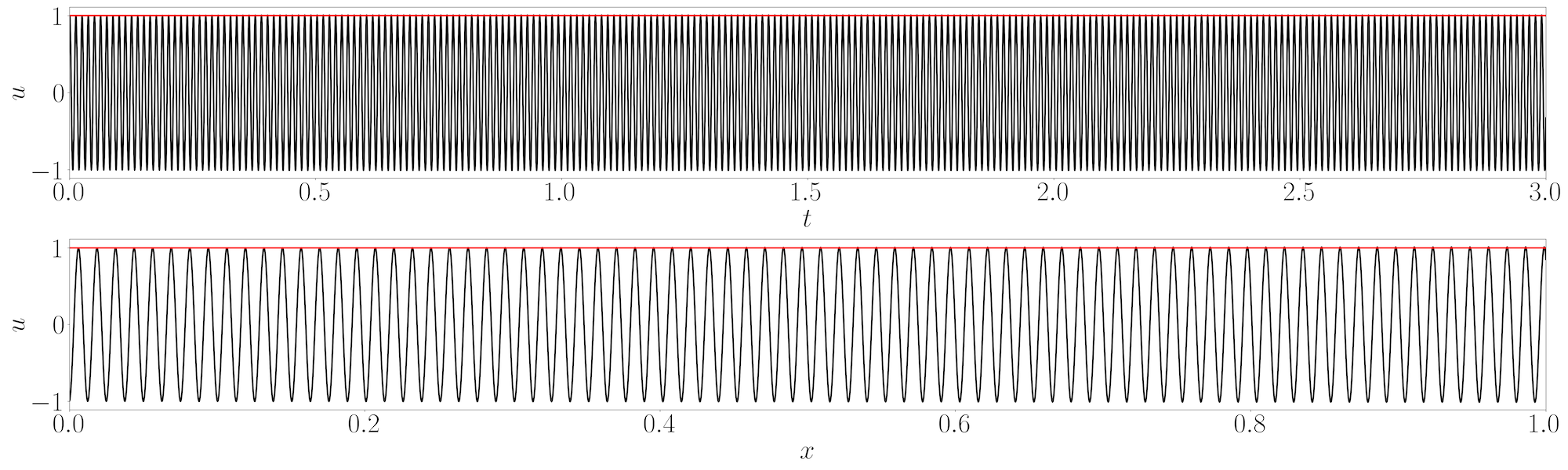}
\caption{Numerical solution of $\omega_{1} = 1.2$ using explicit Euler time integration scheme with $\Delta t = 1.0 \times 10^{-7}$. Top, probed solution at $x=0.5$; Bottom, numerical solution at the end of the simulation. The red horizontal line indicates the iso-line $u=1$.}
\label{fig:w0_very_small}
\end{figure}

\begin{figure}[t]
\centering
\includegraphics[width=0.9\textwidth]{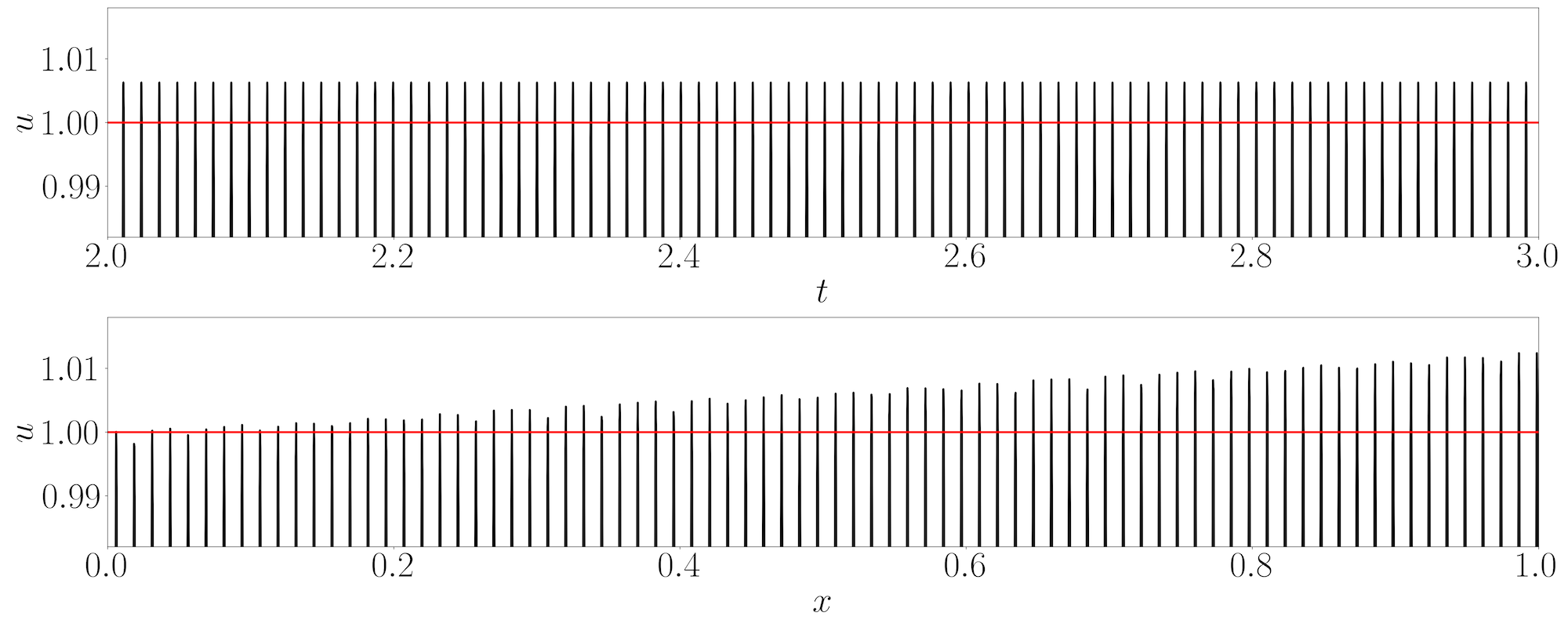}
\caption{Closer look of the solution of fig.~\ref{fig:w0_very_small} around $u=1$ (red line).}
\label{fig:w0_very_small_detail}
\end{figure}

The same simulation has been performed using an even smaller time step $\Delta t = 1.0 \times 10^{-7} = 5 \times 10^{-4} h/(N+1)$. The probed and final solution of the numerical simulation is shown in fig.~\ref{fig:w0_very_small}. Even if barely perceptible, the solution tends to grow while advected along $x$. The same detailed view of fig.~\ref{fig:w0_small_detail} is reproduced in fig.~\ref{fig:w0_very_small_detail}. In this zoomed visualisation, it is possible to observe more clearly the increase in  amplitude of the sinusoidal signal while propagating into the domain, whereas no growth is observed monitoring the probed solution.


The above results confirm that the explicit Euler scheme is marginally stable for pure advection, a well-known deficiency of this method \cite{lomax2013fundamentals}. Still, for under-resolved problems (i.e.\ large frequencies), the fully-discrete scheme revealed to be stable, probably due to the increased upwind dissipation acting against the anti-dissipation of the explicit Euler method. In either case, once the solution establishes itself in space, it does not grow over time --- in line with what is expected within the spatial analysis framework: either stability or convective instability.

\subsection{Accurately resolved flows}
In the theoretical analysis, presented in the first part of the paper, the discussion was organized making distinction between high and low frequencies, considering more or less extended upper bounds for the frequency $\omega$. In the low frequency regions, the reduced order of accuracy for time steps approaching the CFL limit was observed. The convergence results already reported in the temporal eigenanalysis by Vermeire et al.~\cite{vermeire2017behaviour} were correctly reproduced within the present fully-discrete spatial eigenanalysis framework (see fig.~\ref{SD_k4_RK33_conv}). Such feature highlights the similarity between temporal and spatial eigenanalysis for well-resolved wavenumbers/frequencies respectively. 
For high frequencies, instead, the fully-discrete spatial eigenanalysis has shown significant variations from the semi-discrete analysis. In the present subsection we will perform a series of numerical experiments on accurately resolved nonlinear inviscid flows in order to validate the theoretical analysis for well-resolved frequencies. A second subsection will instead be dedicated to under-resolved flows which are commonly associated to high-frequencies which are only partially resolved by the numerical scheme. 
In particular, as a representative example of the spatial eigenanalysis, a variation of the classic isentropic vortex simulation~\cite{shu1998essentially} have been considered. Such formulation is based on an inflow-type condition and  is characterised by spatial grid inhomogeneity due to a sudden change of mesh resolution in the middle of the domain. Clearly, such configuration is particularly representative of spatial eigenanalysis, where an oscillatory flow field is imposed at the inlet boundary and let evolve along the streamwise direction. Finally, in terms of turbulent flows, the study of highly accurate fields, where all the scales and frequencies are sufficiently well-resolved by the numerical scheme, is directly linked to the Direct Numerical Simulation (DNS) approach.

\subsubsection{Inlet-outlet advection of isentropic Euler vortex}
A similar numerical set-up with respect to the classical isentropic vortex test case has been considered in order to avoid the use periodic boundary conditions, which are classically associated to temporal eigenanalysis. A formulation considering inlet/outlet boundaries is, instead, a much more representative example of spatial eigenanalysis, where a temporal frequency is imposed to the system at the inlet and the evolution of the numerical solution is studied while propagating through the space.

At the initial time, two vortices are centered at $x=1/4 L$ and $x=3/4L$, where $L=40$ denotes the full horizontal length of the domain.
Each vortex is characterised by the smooth solution:
\begin{equation}
\begin{aligned}
\rho & = \bigg(1 - \frac{\beta^{2}(\gamma-1)}{16 \gamma \pi^{2}} e^{2(1-r^{2})} \bigg)^{\frac{1}{\gamma-1}}, \\
u & = u_{0} - \frac{\beta (y-y_{c})}{2\pi} e^{1-r^{2}}, \\
v & = v_{0} + \frac{\beta (x-x_{c})}{2\pi} e^{1-r^{2}}, \\
p & = \rho^{\gamma}, \\
\end{aligned}
\end{equation}
where  $r=\sqrt{(x-x_{c})^{2}+(y-y_{c})^{2}}$, with $x_{c}$ and $y_{c}$ indicating the coordinates of the vortex's center, and $\gamma=1.4$ indicates the specific heat ratio. The parameter $\beta$ defines the strength of the vortex and, for this particular study, a value equal to $5$ has been chosen, providing a relatively strong vortex. 
Subsequently, the solution is advected by the constant velocity field $(u_{0},v_{0}) = (5,0)$ while a constant stream of new isentropic vortices is imposed on the left boundary. After a full period has passed, two new vortices have entered the domain from the inlet boundary, whereas the two initial ones have left the domain through the outlet boundary. 

\begin{figure}[t]
\centering
\includegraphics[width=0.56\textwidth]{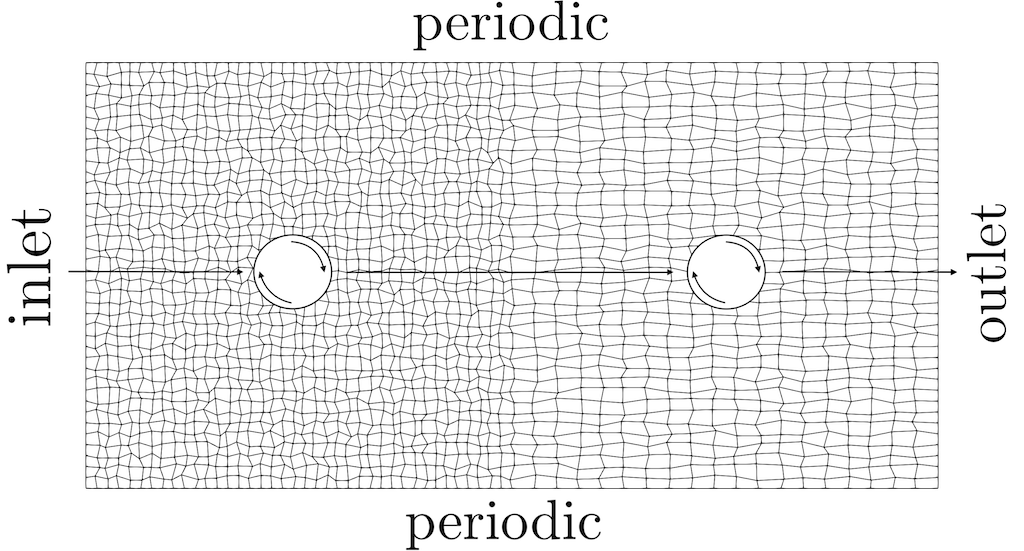}
\caption{Example of deformed computational grid used in the isentropic vortex simulation.}
\label{mesh_ex_spat}
\end{figure}

The smooth analytical solution of the isentropic vortex test case provides an ideal set-up to evaluate the convergence of numerical schemes for the discretisation of Euler equations. The numerical solution of Euler equations has been considered on a series of increasingly refined grids. Namely, the number of elements of the different meshes reads $N_{\mathrm{el}}=N_{x} \times N_{y}=2N_{y} \times N_{y}$ with $N_{y}=10,20,40,80,160$. In order to highlight numerical errors, both regular and deformed grids have been considered. An example of deformed mesh is shown in fig.~\ref{mesh_ex_spat}.
Furthermore, in order to study the evolution of the initial solution while propagating in space, a sudden change on streamwise grid resolution has been placed in the middle of the domain (see fig.~\ref{mesh_ex_spat}). This way, the three-fold influence of inlet, outlet and mesh coarsening can be probed in a single test.
Finally, in order to keep small the numerical errors associated to the spatial discretisation, relatively high orders of approximation have been considered. In particular $6^{\mathrm{th}}$ and $8^{\mathrm{th}}$ order accurate Spectral Difference simulations have been performed, coupled with a $3^{\mathrm{rd}}$ order Runge-Kutta scheme as time integration scheme. In order to evaluate the influence of temporal numerical errors, the time step has been chosen such that $\tau=0.1,0.5,0.9$.

The goal of the present test case consists in showing the essential equivalence between spatial and temporal approaches in the well-resolved regions (the former for well-resolved wavenumbers, the latter for well-resolved frequencies). When we prescribe the stream of vortices through the inlet condition, given the large number of degrees of freedom (DOF) employed in the tests (high polynomial order), we are in fact working in the low-frequency, well-resolved region of spatial analysis. The $L^{2}$ error of the density field is easily evaluated and a convergence study can be performed similarly to the classical test case (whose results are omitted for the sake of brevity).

\begin{figure}[t]
\centering
\subfigure[N=5]{
\includegraphics[width=0.36\textwidth]{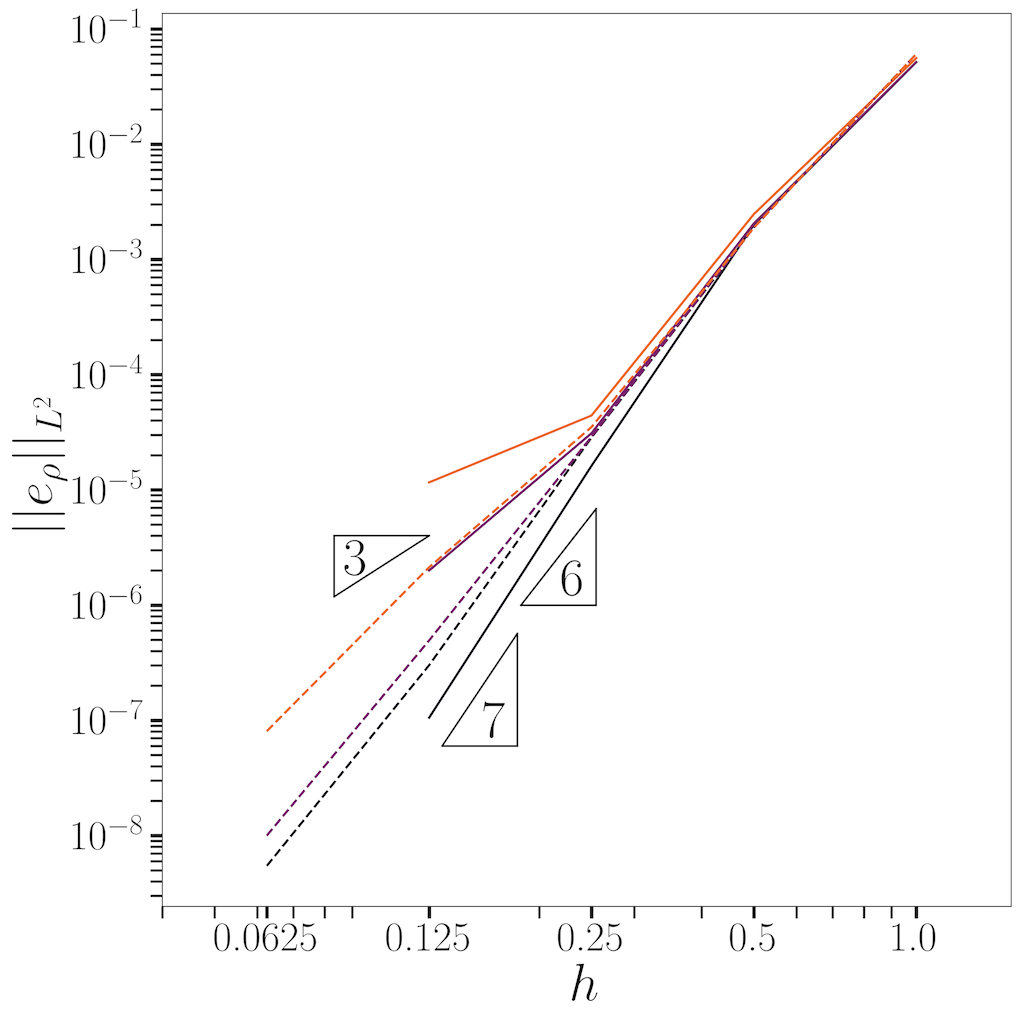}}
\subfigure[N=7]{
\includegraphics[width=0.36\textwidth]{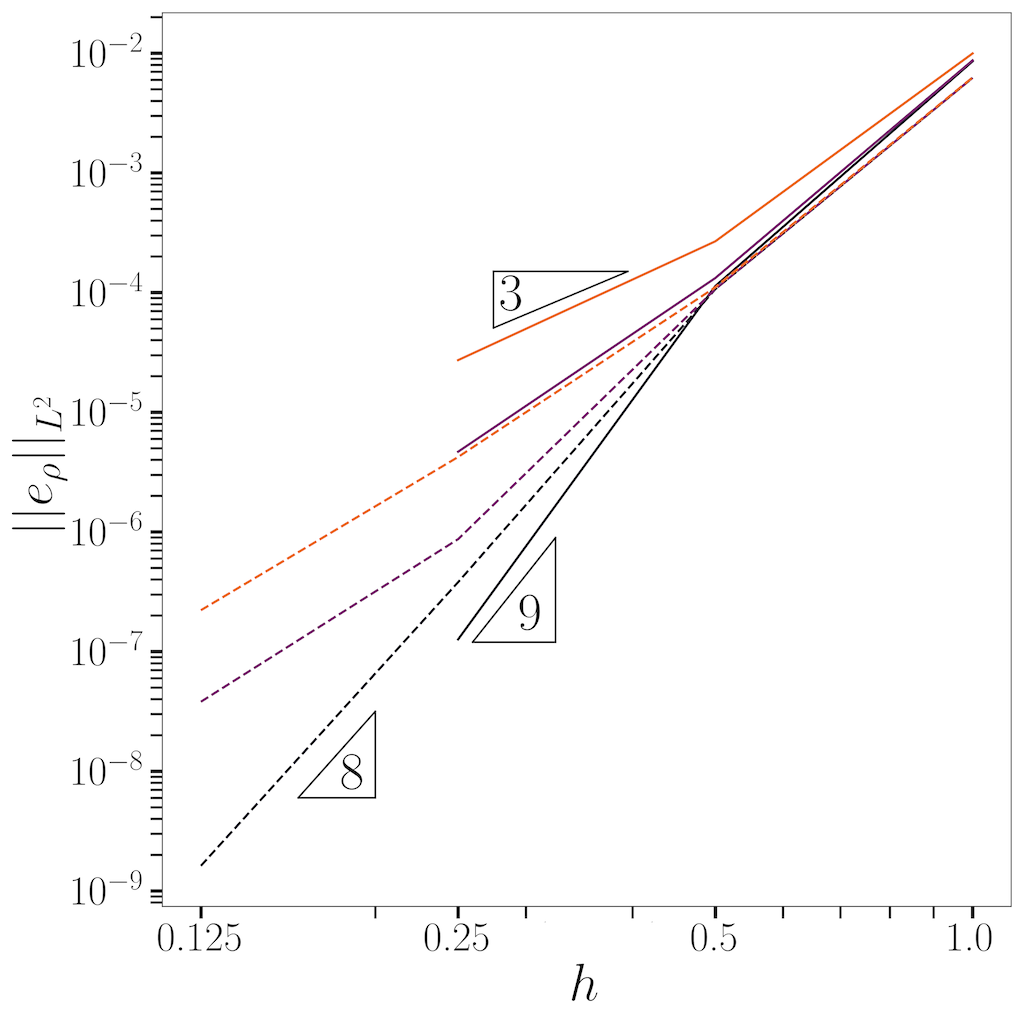}}
\caption{Density's convergence when using RK33. Black line, $\tau=0.1$; purple line $\tau=0.5$; orange line $\tau=0.9$. Solid line, structured grid; dashed line, deformed grid.}
\label{SD_RK3_Euler_conv_spat}
\end{figure}

The convergence plots for the $6^{\mathrm{th}}$ and $8^{\mathrm{th}}$ order simulations in conjunction with RK33 are shown in fig.~\ref{SD_RK3_Euler_conv_spat}. These plots very closely resemble those found for the classical (periodic) case. We note that time steps close to the CFL limit slow down the error convergence. In particular, the expected convergence rate arising from the spatial discretisation is eventually reduced to the order of the time integration scheme. When meshes of stronger deformation are considered, the same behaviour is observed, but at smaller grid sizes. This is believed to happen because of the augmented spatial numerical errors, which can hide the effect of large time steps.

\begin{figure}[t]
\centering
\subfigure{
\includegraphics[width=0.56\textwidth]{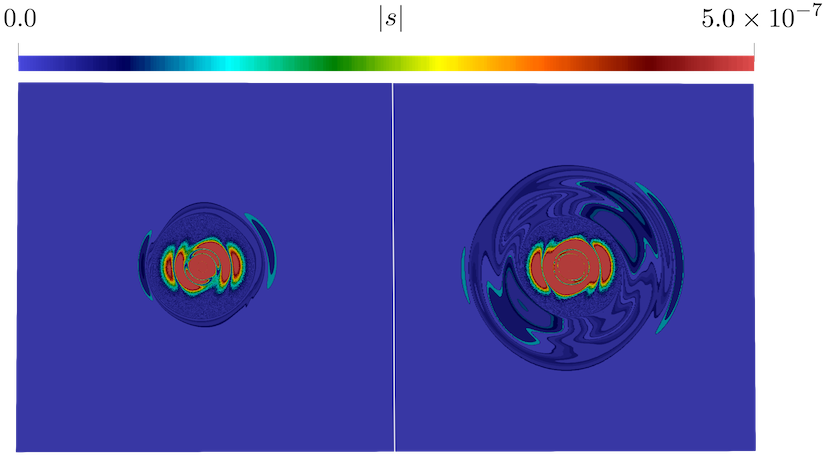}}
\subfigure{
\includegraphics[width=0.56\textwidth]{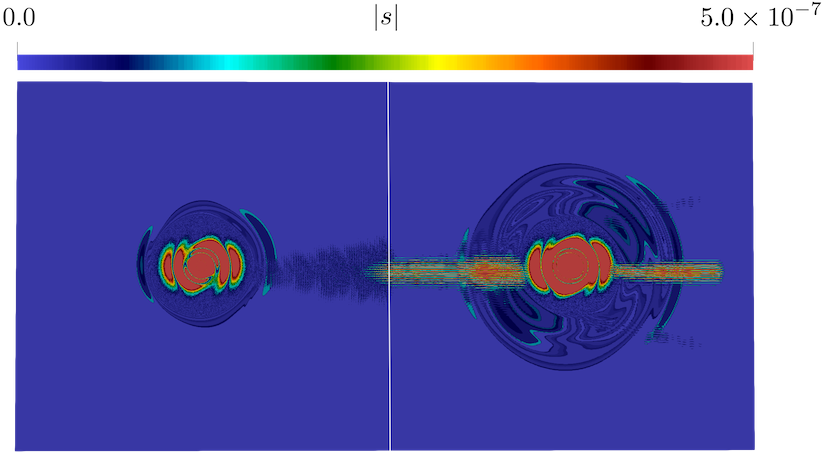}}
\caption{Absolute value of the thermodynamic entropy $s=\ln(\rho^{-\gamma} p)$ throughout the domain for the $6^{\mathrm{th}}$ order simulation coupled with RK33 using $\tau=0.9$ (top) and $\tau=0.99$ (bottom). Vertical white line denotes the location of the mesh coarsening.}
\label{entr09}
\end{figure}

In order to illustrate the influence of the time step on the evolution of the vortex, the absolute value of the thermodynamic entropy $s=\log(\rho^{-\gamma} p)$ has been employed as an indicator of numerical errors. Of course, since an inviscid, shock-free, Euler flow is being considered, the entropy should be theoretically conserved everywhere in the domain. Any deviation from a constant entropy state is then entirely associated to numerical artifacts. The entropy fields at the final time using $\tau=0.9$ and $\tau=0.99$ are shown in fig.~\ref{entr09}, for the $6^{\mathrm{th}}$ order discretisation coupled with RK33. It is interesting to notice that in most of the domain the entropy fields of the two simulations are quite similar. The only relevant differences are represented by the highly irregular spurious oscillations close to the line $y=0$. These are stronger in the second part of the domain, as they are connected with spurious transmitted waves (only barely dissipated as $\tau \rightarrow 1$). Smaller spurious waves are reflected into the first part of the domain too, but are more quickly dissipated as they propagate backwards. In agreement with the work by Vermeire et al.~\cite{vermeire2017behaviour}, a non-negligible influence of the time step choice is observed only very close to the CFL limit. 
%

\subsection{Under-resolved flows}
As mentioned in the previous section, a first comparative study between spatial and temporal eigenanalysis can be performed considering well-resolved flow fields. In such conditions, both theories should return the expected decrease of spatial accuracy for large time steps close to the CFL limit. In the previous section such behaviour has been numerically confirmed. In this second test, instead, high frequencies and wavenumbers (i.e.\ under-resolved flow fields) will be considered. In contrast to the previous section, where well-resolved fields have been naturally associated to DNS, now the behaviour of the numerical scheme represents Large Eddy Simulations (LES, and in particular Implicit LES), where only the large-scale dynamics is adequately resolved by the number of DOF employed. According to the theoretical analysis, for cases employing standard upwind fluxes, large time step values should considerably decrease the overall level of numerical dissipation, affecting the simulation both in terms of stability and accuracy. To perform these tests while focusing on spatial analysis errors, a two-dimensional duct flow has been considered.

\subsubsection{Two-dimensional grid turbulence}
As a typical sample of non-periodic test case, a simple duct flow with oscillating inlet velocity has been considered. The domain consists in a $20\pi \times 2 \pi$ rectangle. Although two-dimensional and inviscid, the present test case is sufficiently insightful to quantify the relevant role of the numerical scheme for under-resolved flows, as already shown in many different works~\cite{mengaldo2018spatial,mengaldo2018spatial2,moura2020spatial,tonicello2021comparative}. 

The inlet boundary condition reads 
\begin{align*}
\rho &= \rho_{\infty}, \\
\vect{u} &= u_{\infty} \begin{pmatrix}
1+A \sin(Ky) \sin(\Omega t) \\
0 \\
\end{pmatrix},\\
p & =p_{\infty},
\end{align*}
where $\rho_{\infty}=1$, $u_{\infty}=1$ and $p_{\infty} = (\rho_{\infty} u_{\infty}^2)/(\gamma \Ma^2)$ in order to get a desired value of inflow Mach number. The parameters defining the inflow perturbations have been set as $A=1/2$, $K=5$, $\Omega=1$. The prescribed Mach number is set equal to $0.03$. For this particular choice, the flow is then essentially incompressible. Non-reflective Roe flux-based boundary conditions are employed at the outlet, whereas top and bottom boundaries are modeled as free-slip walls. The results have been obtained with a $6^{\rm{th}}$-order SD scheme coupled with RK33.

\begin{figure}[t]
\centering
\includegraphics[width=0.99\textwidth]{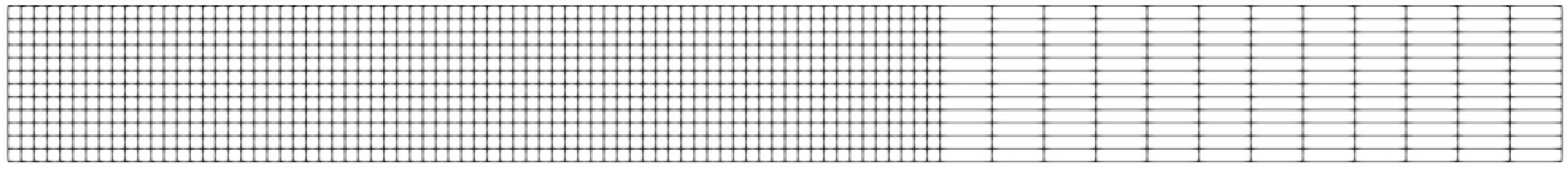}
\caption{Example of computational grid used in the duct flow simulation.}
\label{fig:mesh_ex}
\end{figure}

In order to trigger spurious numerical oscillations, a sudden change in streamwise mesh spacing is imposed at $x=12\pi$, as shown in fig.~\ref{fig:mesh_ex}. The number of elements along the cross-flow direction is $N_{\rm el}^{y}=12$. Along the streamwise direction, this number is $N^{x_{1}}_{\rm el} = 72$ in the first block and $N^{x_{2}}_{\rm el}=12$ in the second block.

\begin{figure}[t]
\centering
\includegraphics[width=0.99\textwidth]{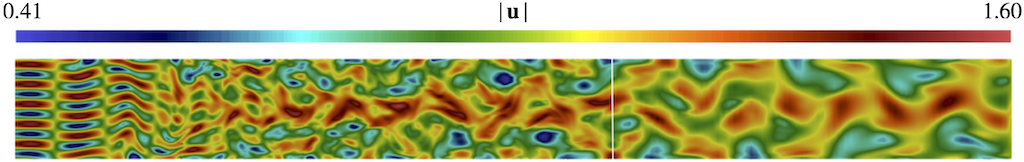}
\caption{Example of duct flow field (velocity magnitude). The white vertical line denotes the change of mesh size interface.}
\label{fig:flow_duct}
\end{figure}

A typical flow field is illustrated in fig.~\ref{fig:flow_duct}, where the two-dimensional nature of the flow leads to a well-known reverse cascade process, with small vortices eventually merging into larger ones. Regardless, the mechanism by which kinetic energy is exchanged across scales is significantly influenced by the underlying numerical errors, whereby the present under-resolved test case is insightful with regards to implicit LES.

Although Roe-type solvers have been found to successfully suppress spurious oscillations coming from the outflow boundary and from the line of mesh size change \cite{mengaldo2018spatial,mengaldo2018spatial2,tonicello2021comparative}, these results typically rely on simulations that are well-resolved in time. The main goal of the present test is to check whether such behaviour is affected by time step size (as predicted by theory) or not. The test case has been considered for $\tau=0.5$, $0.9$ and $0.99$.

\begin{figure}[t]
\centering
\includegraphics[width=0.8\textwidth]{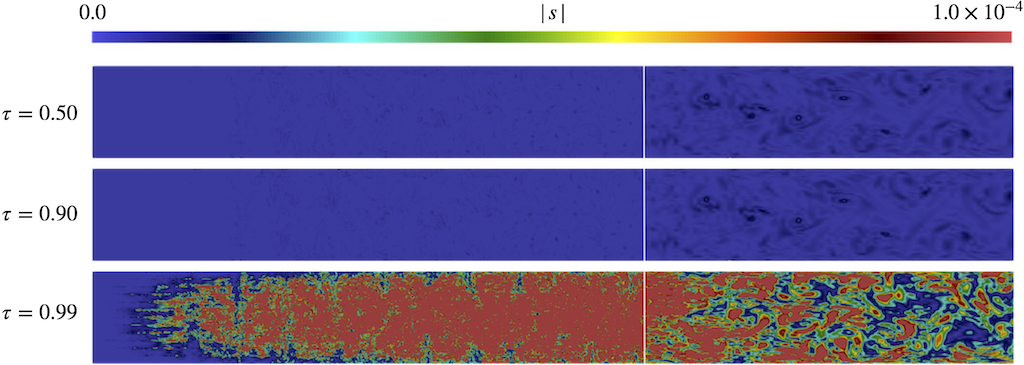}
\caption{Entropy field. The white vertical line denotes the change of mesh size interface.}
\label{fig:entr_duct}
\end{figure}

Similar to the previous test case (isentropic vortex), the time step influence is very small from a \emph{macroscopic} point of view. All cases computed with $\tau < 0.95$ (approximately) behaved quite similarly. Only in very close proximity of the CFL limit, spurious oscillations arose from the numerical solution. As in the previous case, such behaviour is qualitatively represented in fig.~\ref{fig:entr_duct} using the thermodynamic entropy field. Again, spurious transmitted waves are seen when $\tau$ is sufficiently close to unity. In addition, the line of mesh size change seems to trigger more spurious reflected waves than the outflow boundary (at least for this test problem).

Only for values of $\tau$ close to unity, spurious entropy oscillations take place. Due to the structured nature of the mesh, spurious oscillations arise diffusively in the whole domain. In fact, a large number of cells reaches critical values of $\Delta t$ simultaneously, producing spurious entropy oscillations. The same conclusions obtained from isentropic vortex case seem to apply here: the spatial under-resolution hides/overshadows the influence of the prescribed time step. The influence of the time integration scheme is then noticeable only very close to instability, whereas almost any time step fulfilling the CFL condition does not impact the overall accuracy of the numerical solution (already limited by the spatial discretisation).

\section{Conclusions}\label{S:6}
A fully-discrete generalisation of the spatial eigenanalysis has been presented and applied to the Spectral Difference (SD) and Discontinuous Galerkin (DG) schemes. A detailed discussion has been presented on spatial dissipative curves and their dependence on the time step size for 1st- to 4th-order explicit Runge-Kutta time integration schemes. The behaviour for both well-resolved and under-resolved frequencies has been considered. In the fully-discrete spatial eigenanalysis, it has been shown that in the low frequency range, in similarity with  the fully-discrete temporal eigenanalysis, the theoretical spatial convergence rate degrades to the local order of accuracy of the time integration scheme. In the high wavenumbers/frequencies regions, instead, temporal and spatial eigenanalyses have significantly different behaviours. Both are influenced by the time step size and in both cases numerical dissipation generally decreases for larger time steps employing upwind fluxes. In the case of almost centered fluxes, the dissipative bubbles, typical of spatial eigenanalysis of high-order methods, tend to be smoother for increasing time step size. Regions that experience negligible numerical dissipation in the semi-discrete analysis are characterised by stronger diffusion in the fully-discrete generalisation.

The SD discretisation of the one-dimensional linear advection equation has been used to validate the theoretical findings. Numerical experiments have shown good agreement with the theory where the time step size has shown to enhance numerical dissipation for certain frequencies and suppressed it for others. Nonlinear simulations using Euler equations have been performed to quantify the interplay between spatial and temporal errors for more complex dynamics. In agreement with the previous discussion, the Euler simulations have been grouped in well-resolved and under-resolved flows. An inlet-outlet formulation of the classical isentropic vortex test case have been used to validate fully-discrete temporal and spatial eigenanalyses for well-resolved flows. As predicted by the theory, the expected convergence rate, for sufficiently large time steps, is significantly polluted by temporal errors.

In the case of under-resolved flows, instead, the influence of the time step size has shown to be negligible. Spurious oscillations become large enough to visibly pollute the solution only for very large time steps approaching the CFL limit. In under-resolved flows, in fact, the spatial errors dominate their temporal counterparts, overshadowing the influence of the time integration scheme. In conclusion, the theoretical framework introduced in the first part of the paper is not always representative of more complex nonlinear dynamics. For well-resolved flows, the slower convergence has been observed in both the theoretical framework and in the Euler flow simulations. Consequently, the choice of the time step size plays a more relevant role in highly accurate simulations such as DNS of turbulent flows. Instead, in the presence of significant spatial under-resolution (like in LES), temporal errors are more likely overshadowed by spatial errors.

\section*{Acknowledgements}
The use of the SD solver originally developed by Antony Jameson's group at Stanford University is gratefully acknowledged.
This work was granted access to the high-performance computing resources of CRIANN.
The PhD scholarship of the first author is founded by the University of Rouen Normandie.
RCM acknowledges support from São Paulo Research Foundation (FAPESP) via grant 2020/10910-8.
GL acknowledges support from the \emph{Agence Nationale de la Recherche} (ANR)  under grant number ANR-18-CE05-0030.
GM acknowledges support from National University of Singapore start-up grant WBS R-265-000-A36-133. 
The authors also thank David Kopriva, Tristan Montoya and David Zingg for fruitful discussions.

\section*{Appendix} \label{S:A}

\begin{figure}[H]
\centering
\includegraphics[trim=0 0 0 0, clip,width=0.98\textwidth]{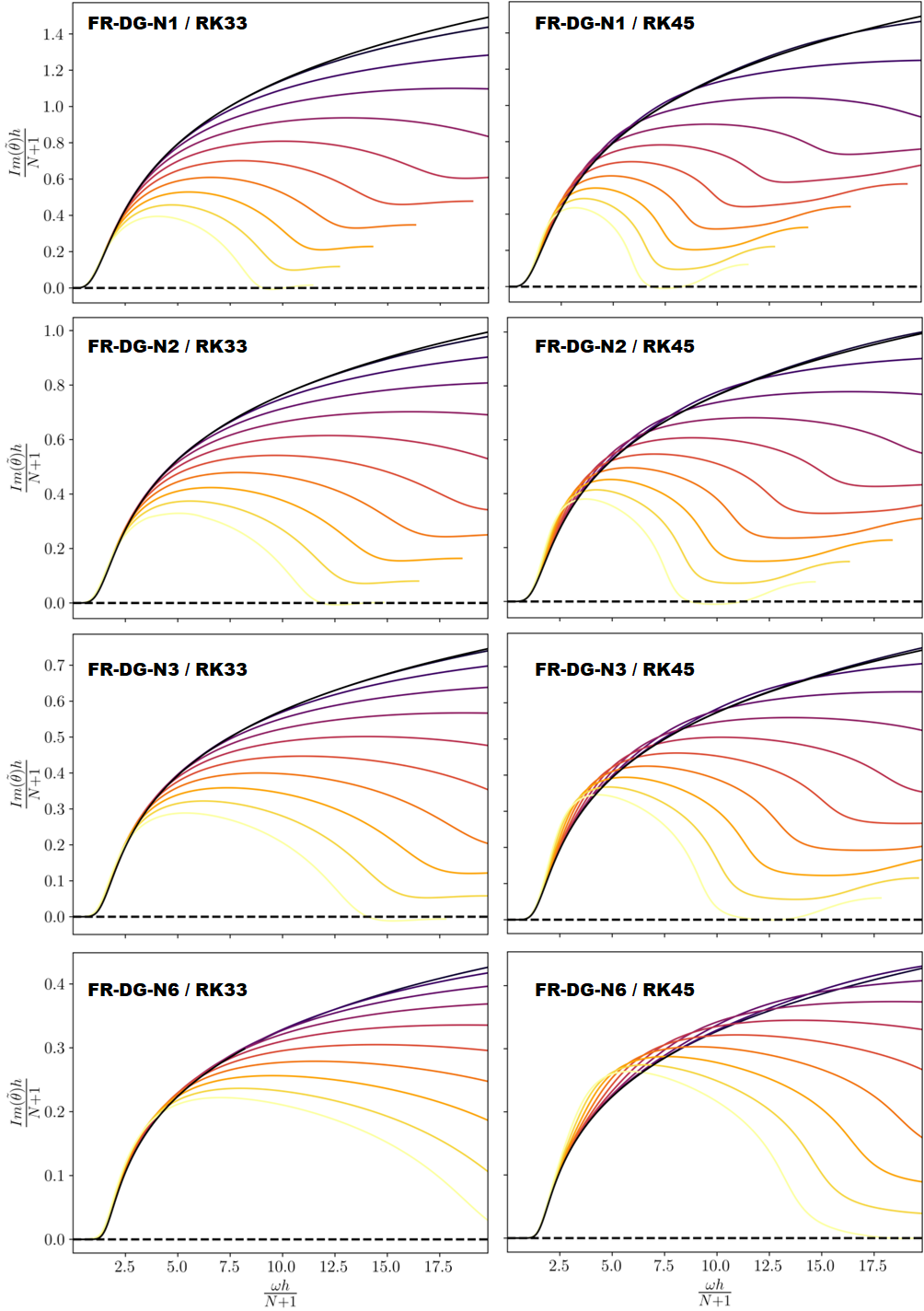}
\caption{Dissipation of the FR-DG scheme for RK33 (left) and RK45 (right) at polynomial orders $N=1,2,3,6$ (top to bottom).}
\label{fig:DGcompile}
\end{figure}

\begin{figure}[H]
\centering
\includegraphics[trim=0 0 0 0, clip,width=0.98\textwidth]{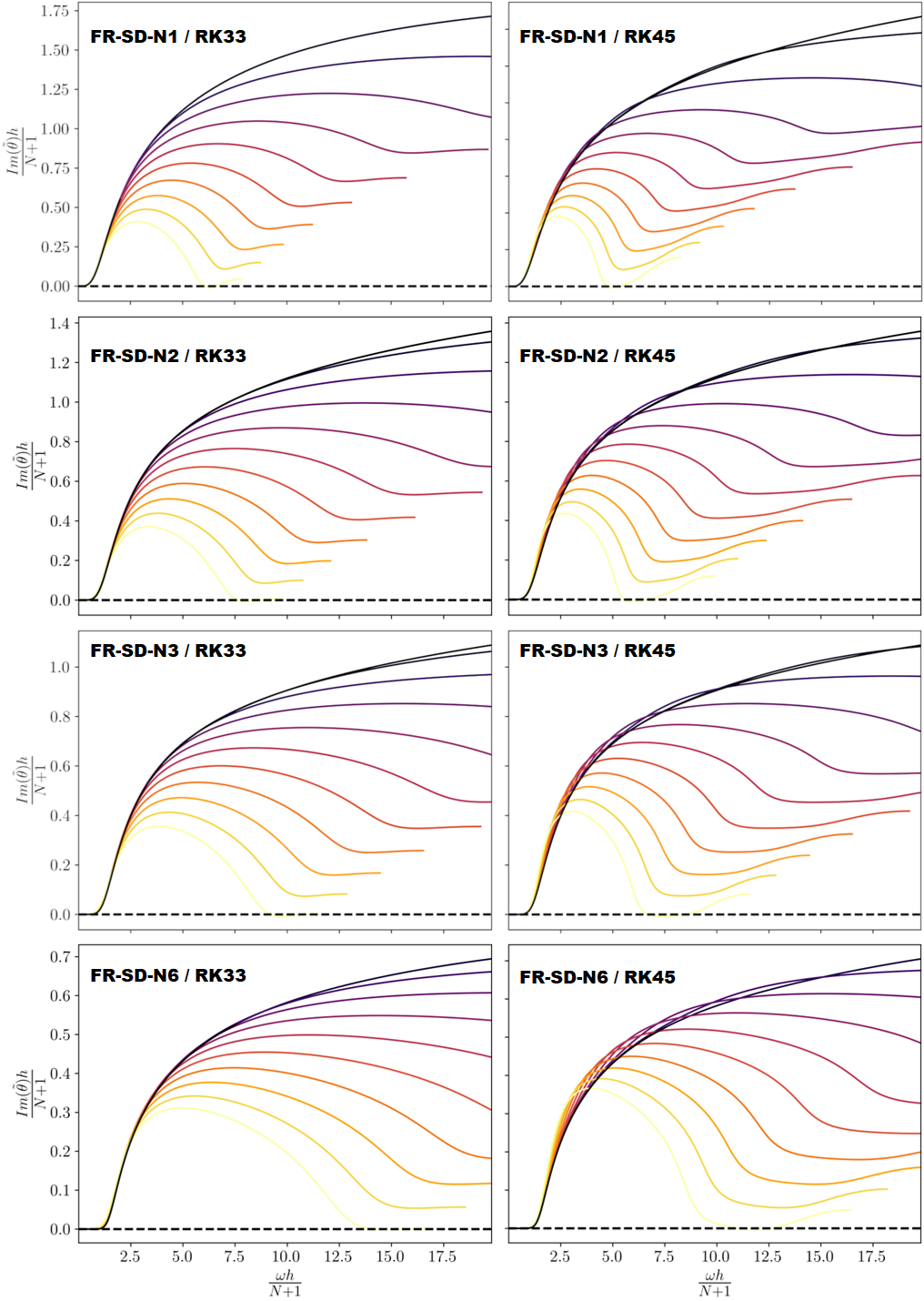}
\caption{Dissipation of the FR-SD scheme for RK33 (left) and RK45 (right) at polynomial orders $N=1,2,3,6$ (top to bottom).}
\label{fig:SDcompile}
\end{figure}

A series of  plots for the dissipation of the FR-DG scheme are given in Fig.~\ref{fig:DGcompile}, whereas  plots for the  FR-SD scheme are given in Fig.~\ref{fig:SDcompile}. The SD curves look similar to those of DG, but are moderately shifted up (stronger dissipation, note the vertical axis) and to the left (toward lower frequencies).

\begin{figure}
\centering
\includegraphics[trim=0 0 0 0,width=0.98\textwidth]{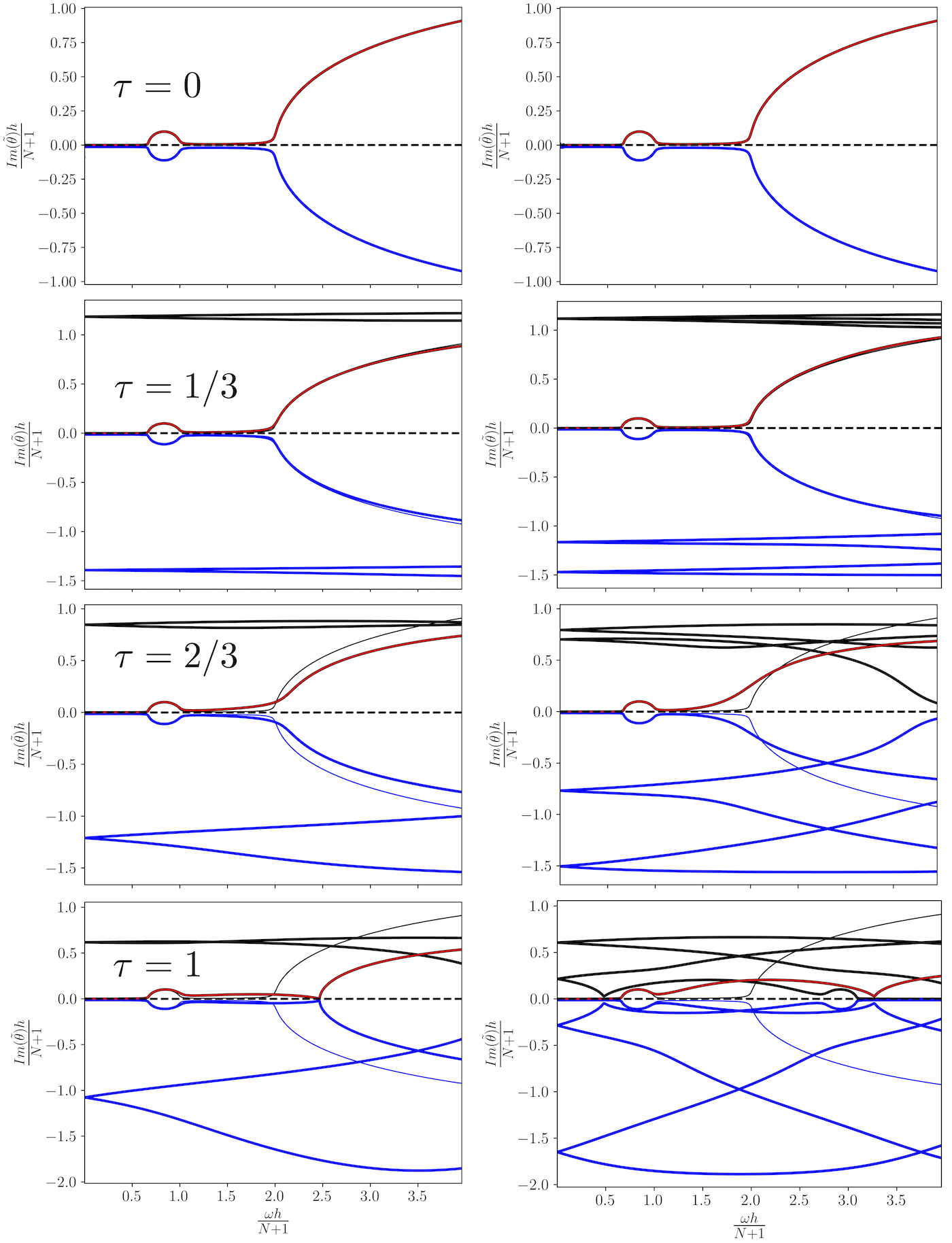}
\caption{Numerical dissipation for a $3^{\rm{rd}}$ order FR-SD discretisation using centred numerical fluxes coupled with RK33 (left) and RK45 (right) time integration schemes. The thinner curves closer to the dashed line represent the semi-discrete result (no temporal error).}
\label{fig:SDk2cent}
\end{figure}

\begin{figure}
\centering
\includegraphics[trim=0 0 0 0, width=0.98\textwidth]{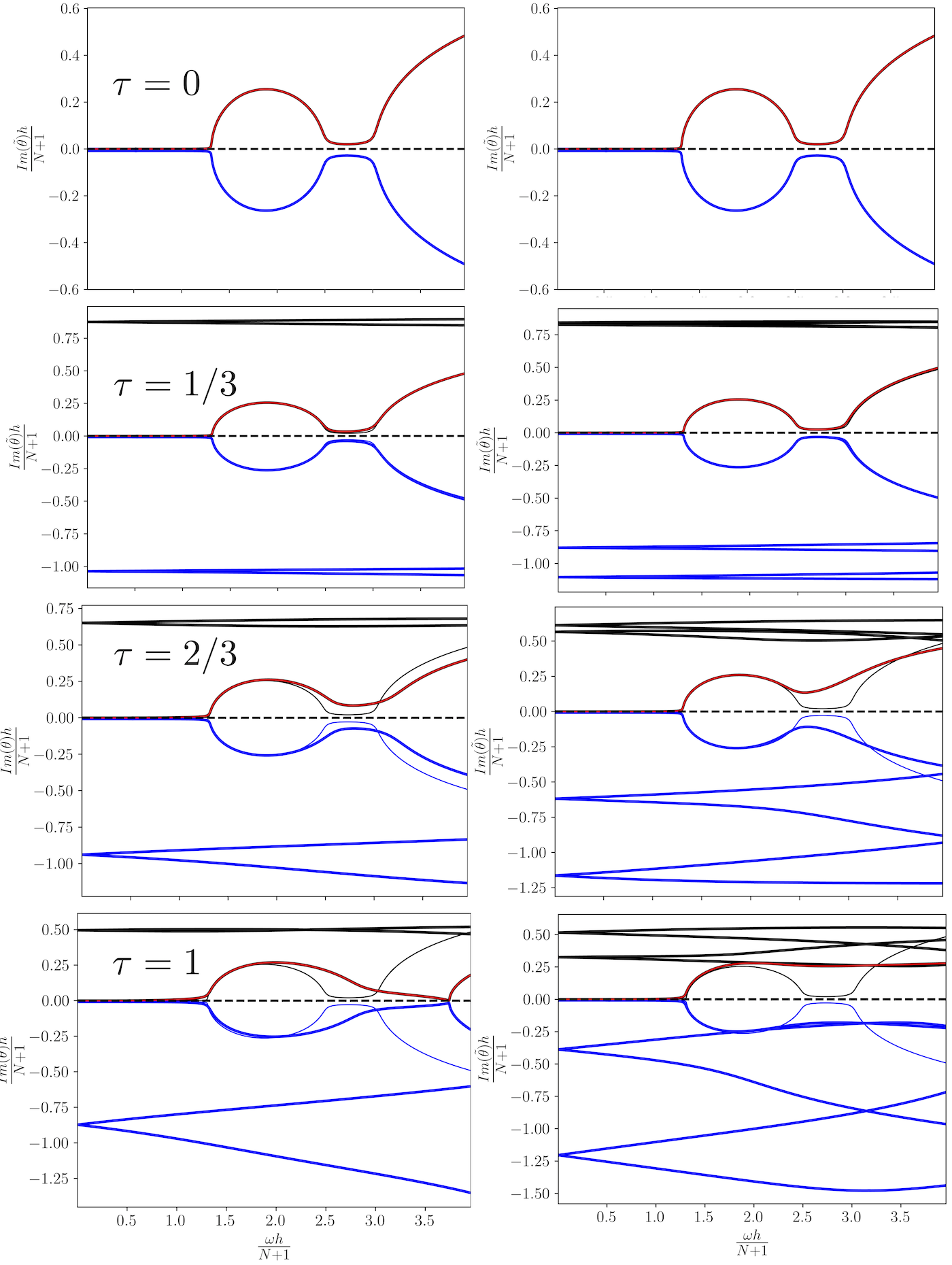}
\caption{Numerical dissipation for a $5^{\rm{th}}$ order FR-SD discretisation using centred numerical fluxes coupled with RK33 (left) and RK45 (right) time integration schemes. The thinner curves closer to the dashed line represent the semi-discrete result (no temporal error).}
\label{fig:SDk4cent}
\end{figure}

In Fig.~\ref{fig:SDk2cent}, the evolution of the multiple modes of the (nearly) centred FR-SD with RK33 and RK45 are shown for $N=2$. Similar plots for $N=4$ are shown in Fig.~\ref{fig:SDk4cent}.
%
%
%

%
%

%
%

\bibliography{JCP_fully_discrete}

\end{document}